\documentclass[a4paper,11pt]{article}
\pdfoutput=2
\usepackage{jheppub} 
\usepackage{epsfig}
\usepackage[caption=false]{subfig}
\usepackage{amsmath}
\usepackage{amsfonts}
\usepackage{amssymb}
\usepackage{float}
\usepackage{color}
\usepackage{graphicx}
\usepackage{graphics}
\usepackage{hyperref}
\usepackage{adjustbox,array}
\hypersetup{}
\usepackage[utf8x]{inputenc} 
\usepackage{epstopdf}
\usepackage{multirow}
\usepackage{slashed}
\usepackage{booktabs}
\usepackage{gensymb}
\usepackage{comment}
\usepackage{diagbox}
\usepackage{autobreak}

\newcommand{\tr}{\mathrm{Tr}}

\hypersetup{
 colorlinks=true,
 linkcolor=blue,
 urlcolor=blue,
 citecolor=red
}

\preprint{CTPU-PTC-23-35, CERN-TH-2023-235, 
\begin{flushright} OU-HET-1217
\end{flushright} }

\title{Dark matter as the trigger of flavor changing neutral current decays of the top quark }

\author[a]{Adil Jueid}
\affiliation[a]{Particle Theory and Cosmology Group, Center for Theoretical Physics of the Universe, \\ 
Institute for Basic Science (IBS), \\
Daejeon, 34126, Republic of Korea}
\emailAdd{adiljueid@ibs.re.kr}

\author[b]{and Shinya Kanemura}
\affiliation[b]{Department of Physics, Osaka University, \\ 
Toyonaka, Osaka 560-0043, Japan}
\emailAdd{kanemu@het.phys.sci.osaka-u.ac.jp}

\begin{document}

\abstract{We suggest a simplified model that simultaneously addresses the dark-matter problem and give rise to top quark flavor changing neutral current (FCNC) interactions at the one-loop order. The model consists of two extra $SU(2)_L$ gauge singlets: a colored mediator of spin zero ($S$) and a right-handed fermion ($\chi$) both are odd under an ad-hoc $Z_2$ symmetry. The right-handed fermion plays the role of the dark-matter candidate. In this model, the presence of the two dark sector particles generates one-loop induced FCNC decays of the top quark into light quarks and bosons such as the gluon, the photon, the $Z$-boson or the Higgs boson. As a case study, we analyze the top quark FCNC decays into light quarks ($u$ or $c$) and a $Z$ or Higgs bosons. We then study the reliable solutions to the dark-matter problem by estimating the regions in the parameter space that are consistent with the \textsc{Planck} measurement of the dark-matter relic density. We also revisit the bounds from the searches of dark matter in events with at least one high-$p_T$ jet and large missing transverse energy at the Large Hadron Collider (LHC). We then define four benchmark points that are consistent with the existing constraints from collider experiments and cosmology. We finally estimate, for these benchmark scenarios, the rates of a broad range of channels that can be used to probe the connection between the top FCNC transitions and dark matter both at the HL-LHC and a future $100$ TeV collider.}


\maketitle

\section{Introduction}

The Standard Model (SM) of elementary particles predicts that tree-level flavor changing neutral currents (FCNC) transition rates are exactly zero, including those of the top quark. The reason for this is that the biunitary transformations that diagonalize the fermion mass matrices lead to diagonal couplings of the Higgs and the $Z$ bosons to fermions. At the one-loop order, the rates of {\it e.g.} top quark FCNC decays are very suppressed thanks to the Glashow-Iliopoulos-Maiani (GIM) mechanism \cite{Glashow:1970gm}. As a consequence of the GIM mechanism, the suppression of the top quark FCNC rates at the one-loop order is due to the unitarity of the Cabibbo-Kobayashi-Maskawa (CKM) matrix and the smallness of the mass splittings between the different quarks running in the loop. The calculation of the top quark FCNC decay rates in the SM has been done nearly thirty three years ago by the authors of ref. \cite{Eilam:1990zc}. It is found that the branching ratios of $t \to c X$ and $t \to u X$ are generally very small ranging from $10^{-17}$ to $10^{-12}$. Such small rates imply that any observation of the top quark FCNC phenomena at the LHC is a clear sign of new physics beyond the SM (BSM). Several BSM models may give rise to sizeable rates for the top quark FCNC decays \cite{Hou:1991un,Atwood:1996vj,Lopez:1997xv,Yang:1997dk,Eilam:2001dh,Aguilar-Saavedra:2004mfd,Gaitan:2004by,Frank:2005vd,Bejar:2006ww,Agashe:2006wa,Cao:2007dk,Baum:2008qm,Agashe:2009di,Gao:2013fxa,Dedes:2014asa,Abbas:2015cua,Botella:2015hoa,Dey:2016cve,Diaz-Furlong:2016ril,Liu:2021crr,Crivellin:2022fdf,Frank:2023fkc}\footnote{In Table \ref{fig:rates:SM:BSM}, we give a summary of the branching ratios of the top quark FCNC decays in the SM and some of the well-known BSM extensions.}. Searches of top quark FCNC interactions have been carried at the LHC by the ATLAS \cite{ATLAS:2015vhj,ATLAS:2015ncl,ATLAS:2017tas,ATLAS:2018zsq,ATLAS:2018xxe,ATLAS:2018jqi,ATLAS:2019mke,ATLAS:2021amo,ATLAS:2022per,ATLAS:2023qzr,ATLAS:2023ujo} and the CMS collaborations \cite{CMS:2013knb,CMS:2016obj,CMS:2017wcz,CMS:2017bhz,CMS:2021gfa}. It is found that bounds on the top quark FCNC branching ratios have been improved by about an order of magnitude. This achievement is due to two important factors: {\it (i)} the increase on both the center-of-mass energy of the $pp$ collisions at the LHC and on the accumulated luminosity, and {\it (ii)} the improvements on the analysis techniques used to perform the signal-to-background optimization, {\it i.e.} going from simple cut-based methods to novel Machine Learning techniques.  It is expected that the HL-LHC with $3000$ fb$^{-1}$ of integrated luminosity would enable for more stringent bounds on the top quark FCNC decays which will therefore put stronger constraints on various BSM scenarios. 

In this study we consider the possibility that dark matter (DM) and top quark FCNC transitions are connected via new particles that are odd under an ad-hoc $Z_2$ symmetry. In this category of models, both the production of DM at hadron colliders and its self-annihilation in the universe are mediated by colored mediators in $t$--channel diagrams. The DM phenomenology at colliders is minimally realized by extending the SM by one mediator ($Y$) that transforms as a triplet under $SU(3)_c$ and a neutral particle ($X$) under $U(1)_Y$. The mediators can transform, however, either as singlets or doublets under $SU(2)_L$ while the DM candidate must always transform as a singlet. Assuming that the colored mediator is a singlet under $SU(2)_L$, there are 12 possible categories for these models depending on the underlying assumptions on the spin of the mediators and of the DM candidate. The phenomenology of this class of models have been extensively studied in the literature \cite{Chang:2013oia,An:2013xka,DiFranzo:2013vra,Ibarra:2015fqa,Garny:2015wea,Ko:2016zxg,Mohan:2019zrk,Arina:2020udz,Arina:2020tuw,Becker:2022iso,Arina:2023msd}. In the most minimal realizations of these scenarios, the main assumption that has been usually used is that each mediator couples {\it solely} to one quark generation. This assumption is motivated by the requirement to avoid one-loop induced FCNC transitions. Here, we assume however that the colored mediator(s) can couple simultaneously to all the quark generations with {\it generally} different coupling parameters. We consider a minimal extension that contains one colored scalar that possesses the same quantum numbers as a right-handed up-type quark and a Majorana fermion both are singlets under $SU(2)_L$. By taking into account this assumption, this model will lead to {\it non-zero} rates for top quark FCNC transitions at the one-loop order. By analyzing the top quark FCNC decays into $qZ$ and $qH$ within this model and studying the possible constraints from collider experiments and cosmology, we find interesting scenarios that may be amenable to discovery both at the HL-LHC and a future $100$ TeV collider. 

The rest of this manuscript is organized as follows. In section \ref{sec:theory} we present the model and its particle content. We discuss in details the branching ratios for top quark FCNC decays into $qZ$ and $qH$ in section \ref{sec:top}. Section \ref{sec:dm} is devoted to an analysis of the DM relic density within this model. The impact of LHC constraints from searches of DM in events with multijets plus missing energy on our model is discussed in section \ref{sec:LHC}. We present four benchmark scenarios and discuss their characteristics in  section \ref{sec:BPs}. We draw our conclusions in section \ref{sec:conclusions}.

\begin{table}[!t]
\setlength\tabcolsep{4pt}\renewcommand{\arraystretch}{1.2}
\begin{center}
\begin{adjustbox}{max width=1.05\textwidth}
\begin{tabular}{l c c c c c c}
Process & SM  & 2HDM (FC) & 2HDM (FV) & MSSM & RPV--MSSM & RS \\
\toprule
${\rm BR}(t\to Z c)$ & $1 \times 10^{-14}$ & $ < 10^{-10}$ & $ < 10^{-6}$ & $< 10^{-7}$ & $ < 10^{-6}$ & $ < 10^{-5}$ \\
${\rm BR}(t\to Z u)$ & $7 \times 10^{-17}$  & $-$ & $-$ & $< 10^{-7}$ & $< 10^{-6}$ & $-$ \\
${\rm BR}(t\to g c)$ & $5 \times 10^{-12}$ & $< 10^{-8}$ & $< 10^{-4}$ & $< 10^{-7}$ & $< 10^{-6}$ & $< 10^{-10}$ \\
${\rm BR}(t\to g u)$ & $4 \times 10^{-14}$ & $-$ & $-$ & $< 10^{-7}$ & $< 10^{-6}$ & $-$ \\
${\rm BR}(t\to \gamma c)$ & $5 \times 10^{-14}$ & $< 10^{-9}$ & $< 10^{-7}$ & $< 10^{-8}$ & $< 10^{-9}$ & $< 10^{-9}$ \\
${\rm BR}(t\to \gamma u)$ & $4 \times 10^{-16}$ & $-$ & $-$ & $< 10^{-8}$ & $< 10^{-9}$ & $-$ \\
${\rm BR}(t\to H c)$ & $3 \times 10^{-15}$ & $< 10^{-5}$ & $< 2 \times 10^{-3}$ & $< 10^{-5}$ & $< 10^{-9}$ & $< 10^{-4}$ \\
${\rm BR}(t\to H u)$ & $2 \times 10^{-17}$ & $-$ & $< 6 \times 10^{-6}$ & $< 10^{-5}$ & $< 10^{-9}$ & $-$ \\
\end{tabular}
\end{adjustbox}
\end{center}
\caption{Summary of the branching ratios for top quark FCNC decays in the SM along with the highest predicted values in several well-known BSM extensions. The results are taken from the 2013 Top Working Group Report \cite{TopQuarkWorkingGroup:2013hxj}.}
\label{fig:rates:SM:BSM}
\end{table}

\section{Theoretical framework}
\label{sec:theory}

In this work, we consider a minimal simplified model with a $t$--channel scalar mediator ($S$) that carries a color charge and a right-handed fermion ($\chi$) that plays the role of the DM candidate. In this framework, the DM particle interacts primarily with SM quarks through a Yukawa-type interaction. In this study, we consider one possible scenario where the scalar mediator couples to right-handed up-type quarks\footnote{Besides the minimal model we consider in this study, there are two minimal classes of models depending on how the scalar mediator transforms under $SU(2)_L$ and on the Hypercharge assignments of the scalar mediator. For instance, the scalar mediator may carry the same quantum numbers as a right-handed down-type quark. In this case, there is no influence on the top quark FCNC decays at the one-loop order but only on the rates of the FCNC decays of the SM Higgs boson. In the other scenario, the scalar mediator belongs to a doublet under $SU(2)_L$ which would therefore impact both the top quark FCNC transitions, Higgs boson decays into $b\bar{s}$ and low-energy $B$-meson FCNC decays.}. In this framework, the new states transform as  
\begin{eqnarray}
S: ~ ({\bf 3}, {\bf 1})_{+2/3}, \quad \chi: ~ ({\bf 1}, {\bf 1})_{0},
\end{eqnarray}
where the numbers refer to their representations under $SU(3)_c \otimes SU(2)_L \otimes U(1)_Y$. On the other hand, both the scalar mediator and the DM candidate are odd under $Z_2$ symmetry while all the SM particles are even. To ensure that the DM particle is stable we also require that $M_\chi \leqslant M_S$. The most general Lagrangian is given by
\begin{eqnarray}
{\cal L} \supset {\cal L}_{\rm S} + {\cal L}_{\chi} - V(S, \Phi),
\end{eqnarray}
where ${\cal L}_{S}$, ${\cal L}_{\chi}$ and $V(S, \Phi)$ refer to the kinetic Lagrangian of the mediator, the Yukawa-type Lagrangian of the DM particle and the scalar potential respectively. The first two Lagrangian terms are given by
\begin{eqnarray}
{\cal L}_S + {\cal L}_\chi \equiv i \bar{\chi} \slashed{\partial} \chi^c + \frac{1}{2} M_\chi \bar{\chi} \chi^c + ({\cal D}_\mu S)^\dagger ({\cal D}^\mu S) + \bigg(Y_q \bar{q}^c_R \chi S + {\rm h.c.} \bigg),
\label{eq:Lag:S}
\end{eqnarray}
where the first and second terms refer to the kinetic energy and mass term of the right-hand fermion, the third term refers to the gauge-invariant kinetic term of the scalar mediator and the last term corresponds to the $S\chi q$ interaction (where a sum over the quark generations is implicit). In equation \ref{eq:Lag:S}, $Y_q, q=u,c,t$ are generation-dependent Yukawa-type couplings and  ${\cal D}_\mu$ is the covariant derivative given by:
$$
{\cal D}_\mu = \partial_\mu - i g_s T^a G_\mu^a - \frac{g_1}{2} Y_S B_\mu,
$$
with $T^a = \lambda_1^a/2$ being the generators of $SU(3)_c$, $Y_S$ is the hypercharge of the scalar mediator, $g_1$ and $g_s$ are the coupling constants of $U(1)_Y$ and $SU(3)_c$ gauge groups respectively. Unless stated explicitly in the text, we assume that the DM Yukawa-type couplings are universal in the sense that $Y_u=Y_c=Y_t$. In the last term of the Lagrangian in equation \ref{eq:Lag:S} one can see that only one scalar mediator couples to all the SM quark generations. Therefore, one can generate top quark FCNC decays at the one-loop order mediated {\it solely} by the dark particles of the model. Assuming {\it CP}--conservation, the most renormalizable and gauge-invariant scalar potential is given by 
\begin{eqnarray}
V(S, \Phi) = - m_{11}^2 |\Phi^\dagger \Phi| + m_{22}^2 |S^\dagger S| + \lambda_1 |\Phi^\dagger \Phi|^2 + \lambda_2 |S^\dagger S|^2 + \lambda_3 |S^\dagger S| |\Phi^\dagger \Phi|,
\end{eqnarray}
here $\Phi = (0, (\upsilon + H)/\sqrt{2})^T$ refers the SM Higgs doublet given in the unitary gauge. All the parameters in the scalar potential are assumed to be real-valued parameters. Note that the quartic coupling $\lambda_2$ does not influence the phenomenology of the model and henceforth will be set to be equal to $1$. On the other hand, $\lambda_3$ is subject to constraints from $H\to gg$ and $H\to \gamma\gamma$ signal strength measurements. The effects of the model parameters on the Higgs decay observables is shown in Appendix \ref{sec:Higgs}. Note that in this model, the contribution to the $\rho$  parameter is exactly zero as shown in Appendix \ref{sec:rho}. We close this section by a discussion of the decays of the colored mediator ($S$) in this model. The partial decay width of $S$ into $\chi q$ is given by
\begin{eqnarray}
    \Gamma(S \to q\chi) &\equiv& \frac{Y_q^2 M_S}{16 \pi} \bigg(1 - \frac{M_\chi^2 + m_q^2}{M_S^2}\bigg) \sqrt{\lambda\bigg(1, \frac{M_\chi^2}{M_S^2}, \frac{m_q^2}{M_S^2}\bigg)}, \nonumber \\
    &\approx& \frac{Y_q^2 M_S}{16 \pi} \bigg(1 - \frac{M_\chi^2}{M_S^2}\bigg)^2, \qquad m_q \ll M_S.
\end{eqnarray}
where $\lambda(x, y, z) \equiv x^2 + y^2 + z^2 - 2 (x y + x z + y z)$ is the K\"allen function. A few comments are in order here. First, for $\Delta \equiv M_S - M_\chi \leq m_t$, the scalar mediator decays {\it solely} to an up quark or a charm quark plus $\chi$ with branching ratios satisfying:
\begin{eqnarray}
    \frac{{\rm BR}(S \to u \chi)}{{\rm BR}(S \to c \chi)} \approx \bigg(\frac{Y_u}{Y_c}\bigg)^2.
\end{eqnarray}
Once $\Delta$ becomes larger than the top quark mass, the decay $S\to t\chi$ opens up with a branching going from a few percents near the threshold and becomes very significant for $M_S \gg m_t$. Note that these features are very important in connection with DM phenomenology and collider studies as we will see in the next sections.

\begin{figure}[!t]
\centering
\includegraphics[width=0.8\linewidth]{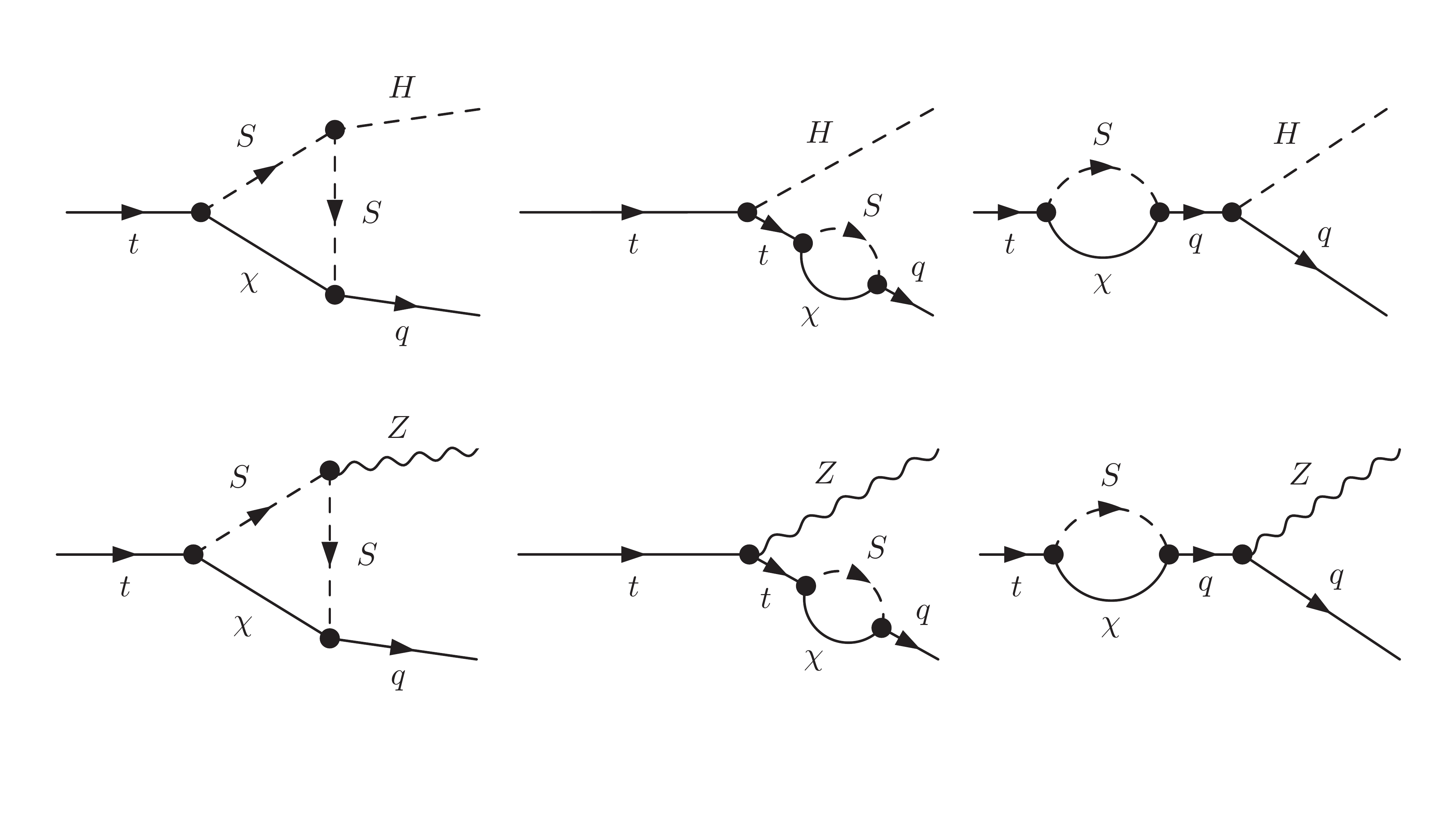}
\vspace{-1cm}
\caption{Examples of Feynman diagrams for the $t\to q H$ (upper panel) and $t\to q Z$ (lower panel) decays.}
\label{fig:FCNC:decay}
\end{figure}

\begin{figure}[!t]
\centering
\includegraphics[width=0.48\linewidth]{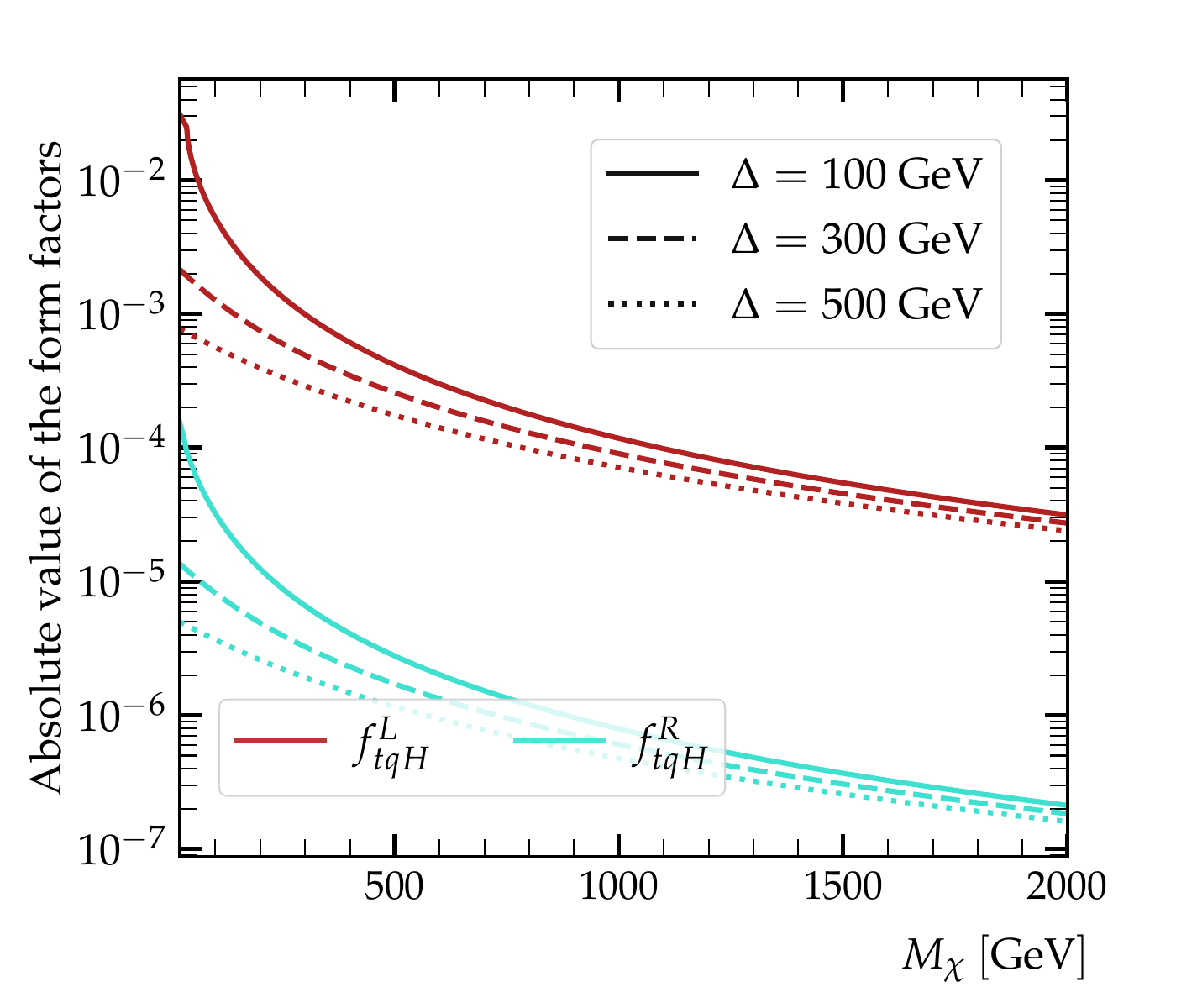}
\hfill
\includegraphics[width=0.48\linewidth]{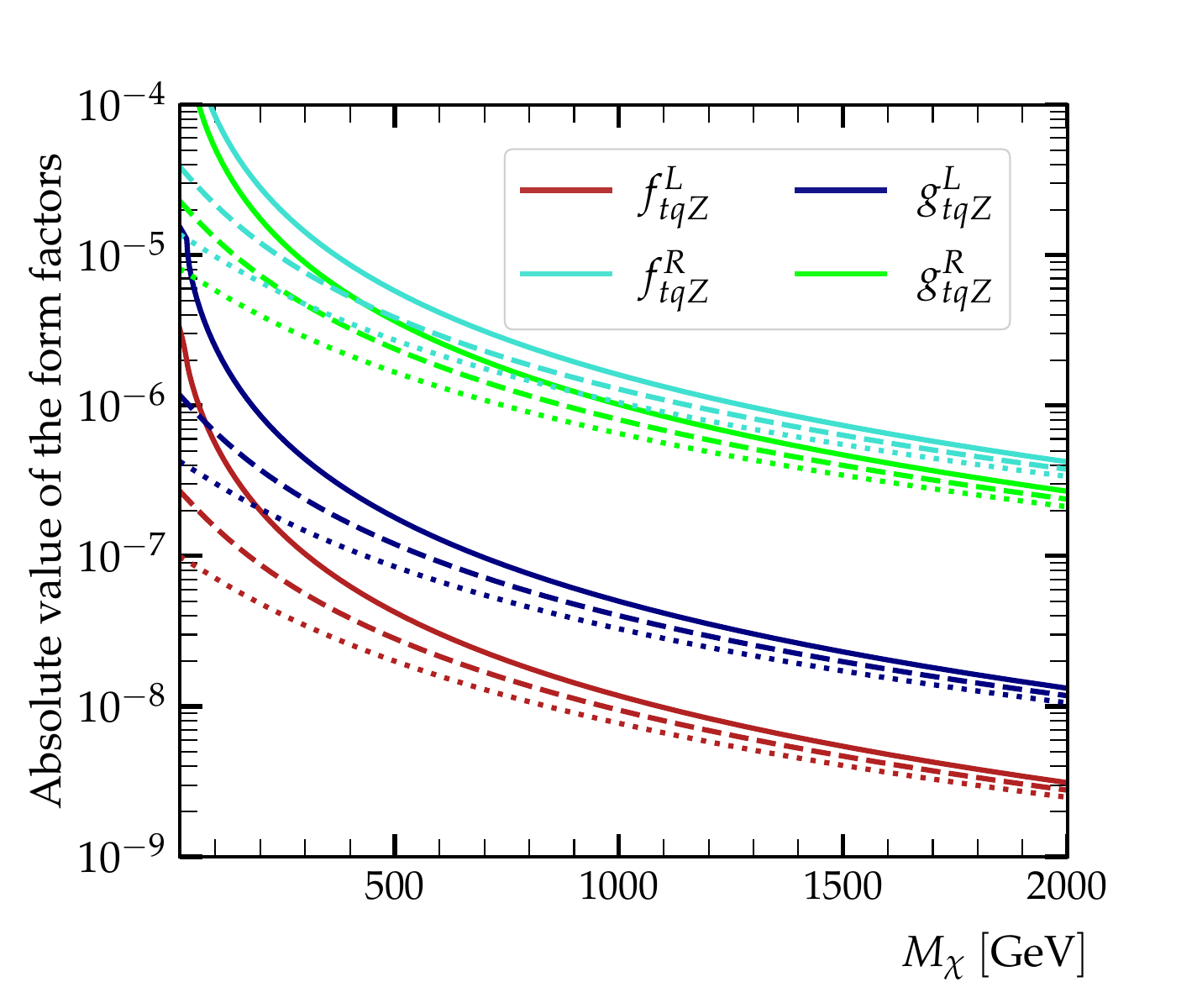}
\caption{The absolute value of the form factors for $t\to q H$ (left) and $t\to q Z$ (right) as a function of $M_\chi$ for $\Delta = 100~{\rm GeV}$ (solid), $\Delta = 300~{\rm GeV}$ (dashed) and $\Delta = 500~{\rm GeV}$ (dashdotted). Here we assume that the light quark $q$ is the charm quark and we take $Y_c = Y_t = 1$ and $\lambda_3 = 1$.}
\label{fig:tqX:ffs}
\end{figure}

\section{Top quark FCNC decays}
\label{sec:top}

In this work, we consider the FCNC two-body decays of the top quark into $q Z$ and $q H$ where $q$ refers to either an up quark or a charm quark. They are mediated by the loops of the scalar mediator and the DM particle as we can clearly see in fig.  \ref{fig:FCNC:decay}. The generic expression of the effective operators for the $t\to q Z$ and $t \to qZ$, $q=u,c$ is given by

\begin{eqnarray}
- {\cal L}_{\rm eff} &=& \bar{t} \gamma^\mu (f_{tqZ}^L P_L + f_{tqZ}^R P_R) q Z_\mu +  \bar{t} p^\mu (g_{tqZ}^L P_L + g_{tqZ}^R P_R) q Z_\mu \nonumber \\
&+& \bar{t} (f_{tqH}^L P_L + f_{tqH}^R P_R) q H + {\rm h.c.},
\end{eqnarray}
where $f_{tqX}^{L,R}~(X=Z,H)$ and $g_{tqZ}^{L,R}$ are form factors that are calculable at the one-loop order, $P_{L,R}=(1\mp \gamma_5)/2$ are the projection operators and $p^\mu$ is the four-momentum vector of the decaying top quark. 

The expressions of the one-loop induced form factors are found using \textsc{FeynArts} version 3.11 \cite{Hahn:2000kx} and \textsc{FormCalc} version 9.9 \cite{Hahn:2006zy} while their numerical evaluations have been done using \textsc{LoopTools} version 2.16 \cite{Hahn:1999mt}. The form factors for $t \to q H$ are given by
\begin{eqnarray}
f_{tqH}^L &=&  \frac{Y_q Y_t m_t}{16 \pi^2} \bigg(3 \lambda_3 \upsilon C_1 + \frac{m_q^2}{\upsilon (m_t^2 - m_q^2)} \left(B_{1,t} - B_{1,q} \right)\bigg), \nonumber \\
f_{tqH}^R &=& \frac{Y_q Y_t m_q}{16 \pi^2} \bigg(3 \lambda_3 \upsilon C_2 + \frac{m_t^2}{\upsilon (m_t^2 - m_q^2)}  \left(B_{1,t} - B_{1,q} \right)\bigg).
\label{eq:tqH:equation}
\end{eqnarray}
While for $t\to qZ$, the form factors are given by
\begin{eqnarray}
f_{tqZ}^L &=& \frac{g_1 m_q m_t (3 c_W^2 - s_W^2)}{96 s_W \pi^2} \frac{Y_q Y_t}{(m_t^2 - m_q^2)} \bigg(B_{1,t} - B_{1,q} \bigg), \nonumber \\ 
f_{tqZ}^R &=& -\frac{g_1 s_W Y_q Y_t}{24 \pi^2} \bigg(2 C_{00} + \frac{1}{m_t^2 - m_q^2} (m_t^2 B_{1,t} - m_q^2 B_{1,q}) \bigg), \nonumber \\ 
g_{tqZ}^L &=& \frac{g_1 s_W Y_q Y_t m_t}{12 \pi^2} \bigg(C_1 + C_{11} + C_{12}  \bigg), \nonumber \\ 
g_{tqZ}^R &=& \frac{g_1 s_W Y_q Y_t m_q}{12 \pi^2} \bigg(C_2 + C_{12} + C_{22}  \bigg).
\label{eq:tqZ:equation}
\end{eqnarray}
In eqs. \ref{eq:tqH:equation} and \ref{eq:tqZ:equation}, $B_{1,Q} \equiv B_{1}(m_Q^2, M_\chi^2, M_S^2)$ and $C_{i,ij} \equiv C_{i,ij}(m_t^2, M_X^2, m_q^2, M_\chi^2, M_S^2, M_S^2)$ refer to the two-point and three-point Passarino-Veltman scalar loop functions \cite{Passarino:1978jh}. Here, $M_X  = m_H$ for $t\to q H$ and $M_X = M_Z$ for $t\to q Z$. It is clear from eqs.  \ref{eq:tqH:equation} and \ref{eq:tqZ:equation} that $f_{tqH}^L \gg f_{tqH}^R$ and $f_{tqZ}^R \simeq g_{tqZ}^L \gg  g_{tqZ}^R > f_{tqZ}^L$ given the mass dependence of these form factors. This can clearly be seen in fig.  \ref{fig:tqX:ffs}, where we show the absolute values of the form factors as a function of the DM mass ($M_\chi$) for three different values of the mass splitting $\Delta \equiv M_S - M_\chi$. We have checked that our results are free of UV divergences and independent of the nonphysical renormalization scale.

Neglecting light quark masses, the resulting decay widths for $t\to q H$ and $t\to qZ$ are given by 
\begin{eqnarray}
\Gamma(t\to q H) &=& \frac{m_t}{32 \pi} \left(1 - \frac{m_H^2}{m_t^2}\right)^2 |f_{tqH}^L|^2, \\
\Gamma(t \to q Z) &=&  \frac{1}{16 \pi m_t} \left(1 - \frac{M_Z^2}{m_t^2}\right) \bigg[\kappa_1~|f_{tcZ}^R|^2 + \kappa_2~|g_{tcZ}^L|^2 - 2~\kappa_3~{\rm Re}(g_{tcZ}^L f_{tcZ}^{R,*}) \bigg], \nonumber
\end{eqnarray}
with $\kappa_1, \kappa_2$ and $\kappa_3$ being functions of $m_t$ and $M_Z$:
\begin{eqnarray*}
\kappa_1 &\equiv& \frac{m_t^4}{2 M_Z^2} \bigg(1 + \frac{m_t^2}{M_Z^2} - \frac{2 M_Z^4}{m_t^4}\bigg), \\
\kappa_2 &\equiv& \frac{m_t^2}{8 M_Z^2} \bigg(1 - \frac{M_Z^2}{m_t^2}\bigg) \times \left(m_t^2 - M_Z^2\right)^2, \quad
\kappa_3 \equiv \frac{m_t}{4 M_Z^2} \left(m_t^2 - M_Z^2\right)^2.
\end{eqnarray*}
Note that the inclusion of the light-quark mass effects would induce a correction that is below $0.1\%$ to the partial widths of the top quark.  The resulting branching ratios are given by
\begin{eqnarray}
{\rm BR}(t \to q X) = \frac{\Gamma(t \to qX)}{\Gamma_t},
\end{eqnarray}
where $\Gamma_t \equiv \Gamma(t\to b W) = 1.32~{\rm GeV}$ calculated at Next-to-Next-to-Leading Order in QCD including the finite mass and width effects and Next-to-Leading Order electroweak corrections \cite{Gao:2012ja}. We note that our model predicts that 
$$
\frac{\Gamma(t\to c X)}{\Gamma(t\to u X)} \approx \left(\frac{Y_c}{Y_u}\right)^2,
$$ 
since the $m_q^2$ corrections to the FCNC partial widths are extremely small. This model predicts a ratio
\begin{eqnarray}
   \frac{\Gamma(t \to q Z)}{\Gamma(t \to q H)} \approx \frac{r}{\lambda_3^2},
\end{eqnarray}
where $r$ is a factor that depends on the DM mass and the mass splitting ($\Delta$) and varies between $5$ and $15$. We find that $r$ is smaller for high $M_\chi$ and higher for small $M_\chi$. 

\begin{figure}[!t]
\centering
\includegraphics[width=0.48\linewidth]{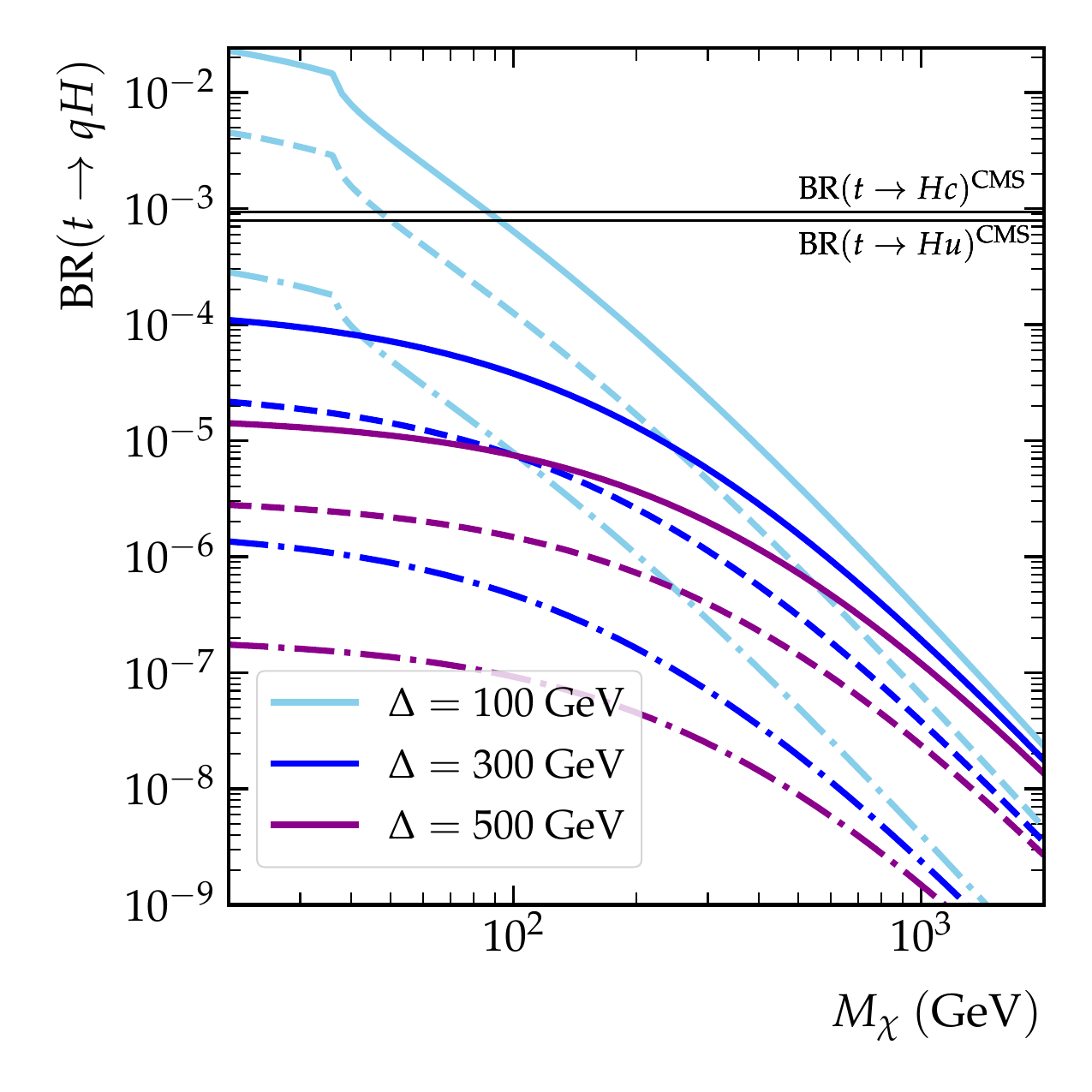}
\hfill
\includegraphics[width=0.48\linewidth]{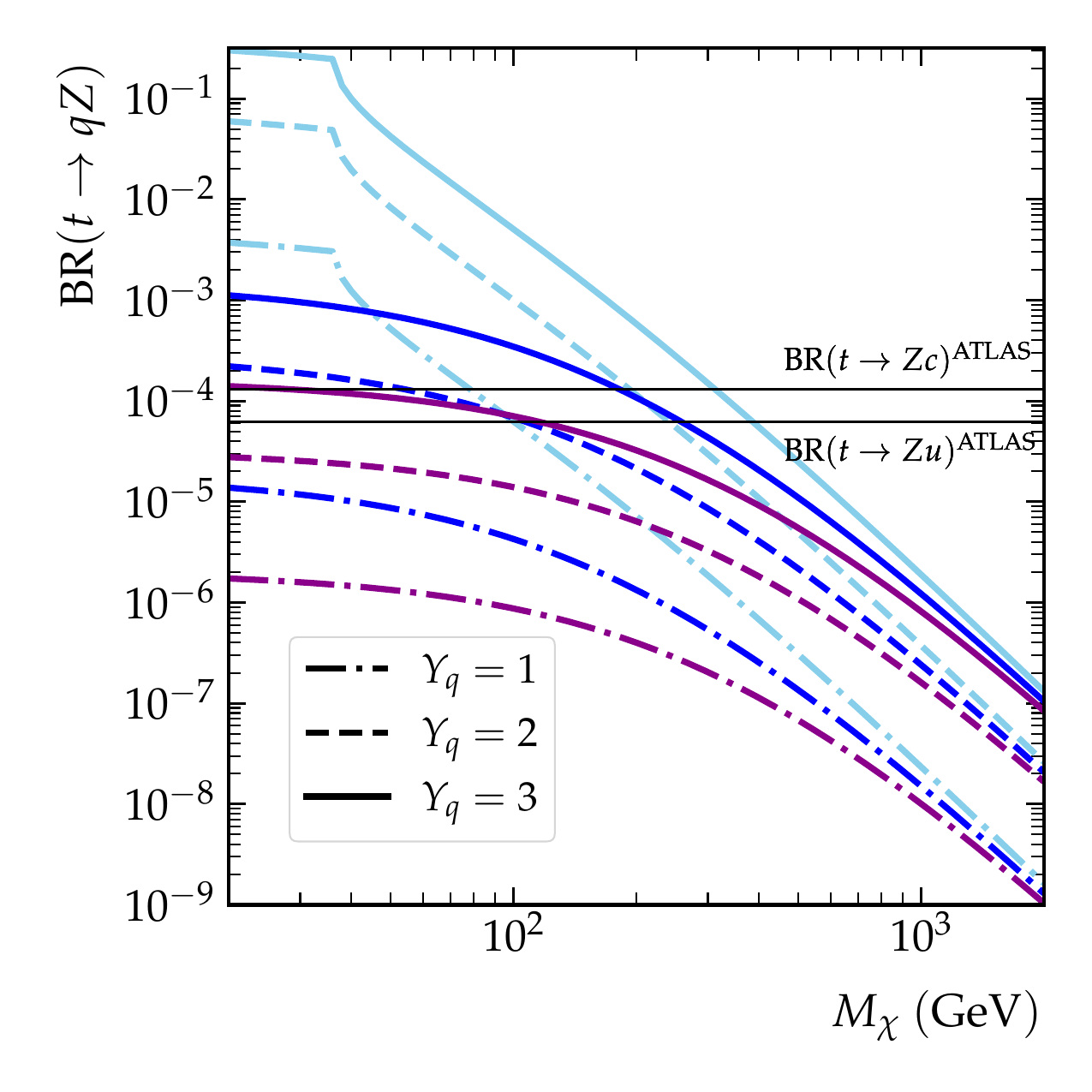}
\caption{The FCNC decay branching ratios as function of the dark matter mass ($M_\chi$) for $t \to q H$ (left) and $t \to q Z$ (right). The results are shown for $Y_q = 1$ (dashdotted), $Y_q = 2$ (dashed) and $Y_q = 3$ (solid). For each case, we calculate the branching ratios for $\Delta = 100~{\rm GeV}$ (turquoise), $\Delta = 300~{\rm GeV}$ (blue) and $\Delta = 500~{\rm GeV}$ (purple). For $t \to q H$, we choose $\lambda_3 = 1$. We also display the latest exclusion bounds from the searches of FCNC production of top quark reported on by ATLAS \cite{ATLAS:2023qzr} and CMS \cite{CMS:2021gfa} collaboration. More details can be found on the text.}
\label{fig:BRtqX}
\end{figure}

In fig.  \ref{fig:BRtqX}, we display the branching ratios of $t\to q H$ (left) and of $t \to q Z$ (right) as a function of $M_\chi$ for different values of the mass splitting $\Delta$ and of the Yukawa-type coupling $Y_q=Y_t$.  Here, we show the results for $Y_q=1, 2$ and $3$ with the choices of $\Delta = 100~{\rm GeV}$ (turquoise), $\Delta = 300~{\rm GeV}$ (blue) and $\Delta = 500~{\rm GeV}$ (purple).  These two FCNC branching ratios scale as $|Y_q Y_t|^2$ and therefore higher values of $Y_q = Y_t$ will lead to extremely large branching ratios especially for small $M_\chi$ and small $\Delta$. We also show the experimental bounds reported on by ATLAS \cite{ATLAS:2023qzr} and CMS \cite{CMS:2021gfa} as horizontal solid (for $t\to u X$) and dashed (for $t \to c X$) lines. The strong bounds from the search of $t\to q Z$ imply that DM masses of order $400~{\rm GeV}$ are excluded at the $95\%$ CL if one assumes that $\Delta = 100~{\rm GeV}$ and $Y_q = 3$. Smaller values of the DM mass are still allowed if one considers $Y_q \approx 1$ and relatively large $\Delta$. On the other hand, we can see that for heavy DM with mass $M_\chi \geq 1000~{\rm GeV}$, the branching ratios become relatively independent of the choice of $\Delta$.  

\section{Dark Matter}
\label{sec:dm}

The relic density of the $\chi$ particles is mainly due to the freeze-out mechanism. The main process that leads to DM relic density within this model is through the 
annihilation into $q_\alpha \bar{q}_\beta$ where $\alpha$ and $\beta$ are generation indices. This includes the case of same-flavor production ($u\bar{u}, c\bar{c}, t\bar{t}$) and of different-flavor production
($u\bar{c}, c\bar{u}, u\bar{t}, t\bar{u}, c\bar{t}, c\bar{u}$). The co-annihilation channels start to dominate for small mass splittings between the $\chi$ particle and the colored mediator $S$. The DM density of the $\chi$ particles can be obtained by solving the following Boltzmann equation assuming that the colored mediator has already decayed into $\chi + q_\alpha$:
\begin{eqnarray}
\frac{{\rm d} n}{{\rm d}t} = -3 H n - \langle \sigma_{\rm eff} \upsilon (x_f) \rangle (n^2 - n_{\rm eq}^2),
\end{eqnarray}
with $H$ being the Hubble parameter, $\langle \sigma_{\rm eff} \upsilon (x_f) \rangle$ is the thermally-averaged annihilation cross section of DM with a velocity $\upsilon$ at the freeze-out temperature $x_f$ and $n_{\rm eq}$ is the equilibrium number density.

\begin{figure}[!t]
\centering
\includegraphics[width=0.49\linewidth]{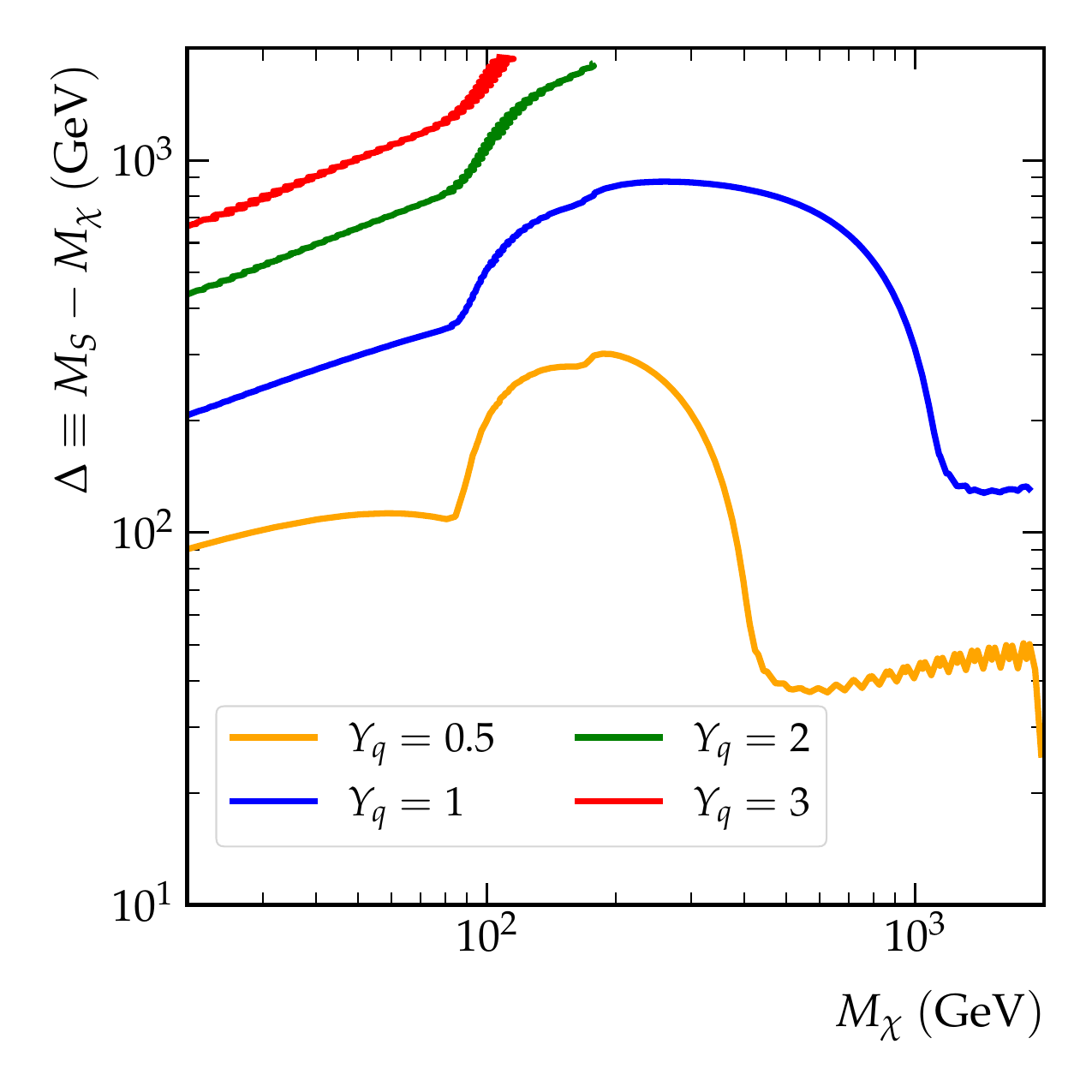}
\hfill 
\includegraphics[width=0.49\linewidth]{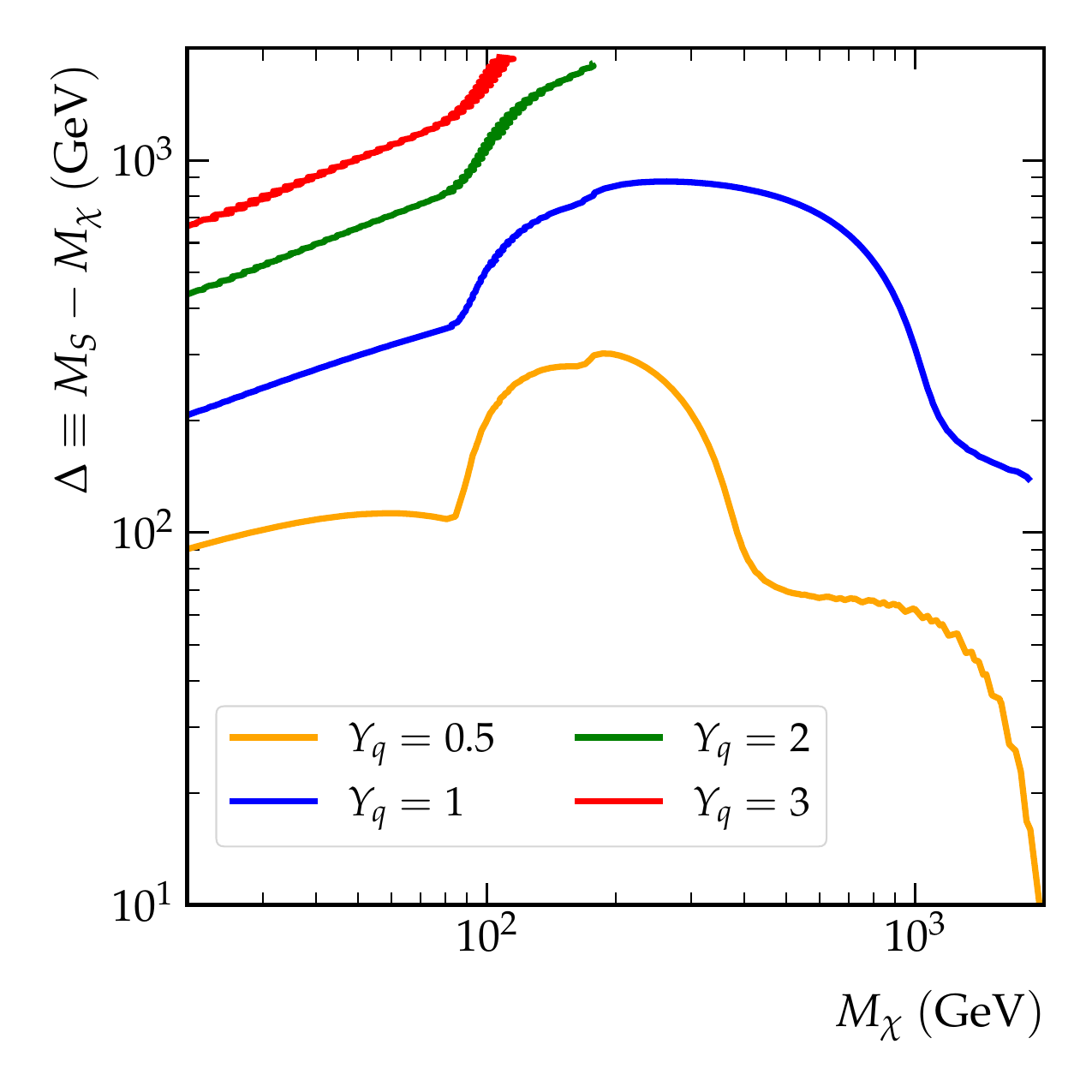}
\caption{Parameter space satisfying $\Omega_\chi h^2 = 0.12$ shown in the $(M_\chi, \Delta)$ plane. The thin solid curves are corresponding to contours of constant coupling $Y_q=0.5$ (orange), $Y_q = 1$ (blue), $Y_q = 2$ (green) and $Y_q=3$ (red) where $Y_q \equiv Y_u = Y_c = Y_t$. Here we show the results without co-annihilation (left) and with co-annihilation (right).}
\label{fig:X:relicdensity}
\end{figure}

An approximate solution of the Boltzmann equation leads to the following expression for the DM relic density \cite{Griest:1990kh,Servant:2002aq,Kong:2005hn}

\begin{eqnarray}
\Omega_\chi h^2 \simeq \frac{1.04 \times 10^{9}}{M_{\rm Planck}} \frac{x_f}{\sqrt{g_{*}(x_f)}} \frac{1}{I_a + 3 \frac{I_b}{x_f}},
\end{eqnarray}
with $M_{\rm Planck} = 1.22 \times 10^{19}~{\rm GeV}$ being the Planck mass, $g_{*}$ is the effective number of degrees of freedom and $I_a$ and $I_b$ are coefficients that depends on the effective cross section. Here, $I_a$ is related to the velocity-independent and $I_b$ is related to the velocity-dependent cross section (see e.g. ref. \cite{Kong:2005hn} for more details). We assume that the thermally-averaged cross section can be expanded as 
\begin{eqnarray}
\sigma_{\rm eff} v = a_{\rm eff} + b_{\rm eff} \upsilon^2  + {\cal O}(\upsilon^4),
\end{eqnarray}
where the expression of $a_{\rm eff}$ and $b_{\rm eff}$ can be found in e.g. ref. \cite{Kong:2005hn}. In this analysis, we use \textsc{MicrOmegas} version 5.3.41 \cite{Belanger:2018ccd} to solve numerically the Boltzmann equations and to calculate the relic density of the $\chi$ particles. We have cross checked the results of \textsc{MicrOmegas} by comparing them with the results of \textsc{MadDM} version 3.1 \cite{Ambrogi:2018jqj} for various benchmark points. The results of the calculations are shown in fig.  \ref{fig:X:relicdensity} where the couplings satisfying the condition $\Omega_\chi h^2 = 0.12$ are shown in the plane of $M_\chi$ and $\Delta = M_S - M_\chi$ in the case where co-annihilation is not included (left) and where co-annihilation is included (right). We show the contours for $Y_q = 0.5$ (orange), $Y_q = 1$ (blue), $Y_q = 2$ (green) and $Y_q = 3$ (red). We can see that in order to fulfill the correct relic density, reasonably modest to large values of $Y_q$ are required which increase with increasing mass splitting $\Delta$. On the other hand, we can see that there is a wide peak for DM masses above $\approx 80$ GeV for a given value of the DM coupling which indicates the opening of the annihilation channel $\chi\chi \to t\bar{t}$. Finally, we note that the effect of co-annihilation becomes very important for small $\Delta$ and large $M_\chi$ (see the right panel of fig.  \ref{fig:X:relicdensity}). \\

We close this section by commenting on the effects of DM direct detection searches on the model parameter space. In this model, the leading contribution to the DM-nucleus spin-independent cross section ($\sigma_{\rm SI}$) occurs through the exchange of the scalar mediator. Further corrections can appear once we consider diagrams at NLO\footnote{The analysis of $\sigma_{\rm SI}$ is very involved and we do not consider it in this work. We refer the interested reader to refs. \cite{Mohan:2019zrk,Belanger:2021smw, Becker:2022iso} for more detailed analyses of direct detection for more generalised models which involve three mediators or one mediator and one scalar leptoquark. We note that these constraints are not strong as long as couplings of order ${\cal O}(1)$ are considered.}. We, however, check the viability of the benchmark points presented in section \ref{sec:BPs} in what concerns the bounds from $\sigma_{\rm SI}$ reported on by the \textsc{Xenon1T} collaboration \cite{XENON:2018voc} using calculations at LO only.

\section{Bounds from the LHC searches}
\label{sec:LHC}

\begin{figure}[!t]
\centering
\vspace{-1.5cm}
\includegraphics[width=0.9\linewidth]{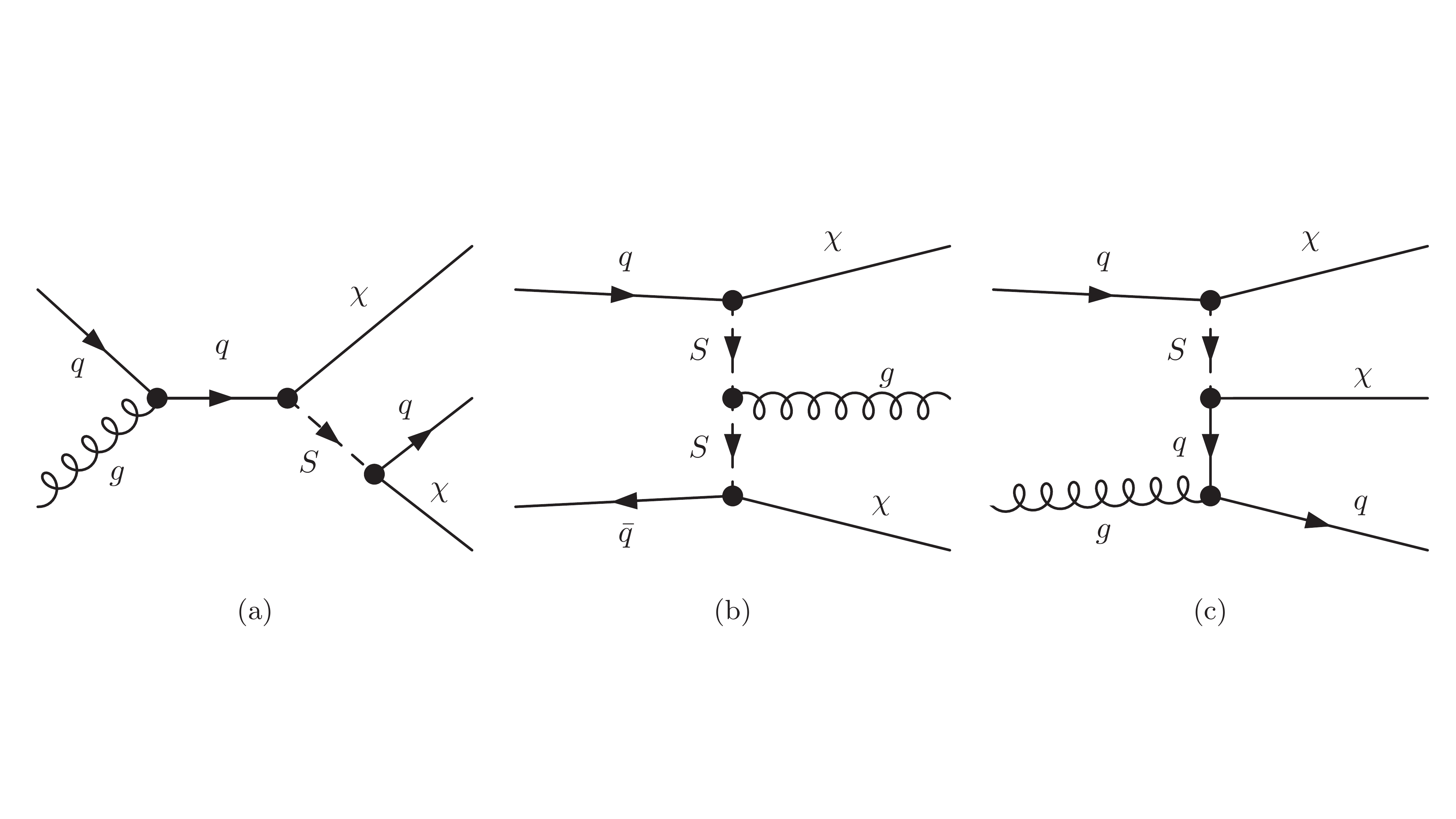}
\vspace{-1cm}
\caption{Representative Feynman diagrams for monojet production in this model. Here, we show the monojet production through resonant $S\chi$ production (a) and noresonant production (b-c).}
\label{fig:FD:monojet}
\end{figure}

The model predicts the production of DM at the LHC through a variety of processes leading to various final-state signatures such as monotop ($t+E_{T}^{\rm miss}$), $t\bar{t}+E_{T}^{\rm miss}$, monojet, and multijet$+E_{T}^{\rm miss}$. There are several collider studies on the models with $t$--channel mediators \cite{Arina:2020udz,Arina:2020tuw,Arina:2023msd}. In these analyses, the scalar mediator is assumed to couple to one generation of quarks only. Therefore constraints from multijet+$E_{T}^{\rm miss}$ are found to be strong excluding a wide range of the parameter space \cite{Arina:2020tuw}. In this model, the fact that the scalar mediator couples to all the three generations of quarks lead to weaker limits. We found that the strongest bound comes from the search of monojet (small number of jets and missing energy) which was carried by the ATLAS collaboration \cite{ATLAS:2021kxv}\footnote{A comprehensive analysis of all the existing collider searches at the LHC will be performed in a future study.}. Examples of Feynman diagrams for mono-jet production in this model at Leading Order (LO) are shown in fig.  \ref{fig:FD:monojet}. In this model, one can distinguish between resonant production (fig.  \ref{fig:FD:monojet}a) and non-resonant production through $q\bar{q}$ annihilation (fig.  \ref{fig:FD:monojet}b) and through $qg$ fusion (fig.  \ref{fig:FD:monojet}c). The cross section of the mono-jet production can be generically expressed as follows:
\begin{eqnarray}
    \sigma(p p \to \chi\chi J) \equiv \sum_{i,j} \int {\rm d}x_i {\rm d}x_j f_{i/p}(x_i,Q^2) f_{j/p}(x_j,Q^2) \hat{\sigma}(ij \to \chi\chi J),
\end{eqnarray}
here $f_{i/p}(x_i,Q^2)$ is the PDF of a parton $i$ within the proton to carry a momentum fraction $x_i$ at a factorization scale $Q$, $\hat{\sigma}$ is the partonic cross section which scales in this process as $\hat{\sigma}(ij\to \chi\chi J) \propto Y_q^4$. Here $Y_q$ is either $Y_u$ or $Y_c$. Note that for all the diagrams, the contribution of the charm quark PDF is always smaller than the contribution of the up or down quarks.  
The analysis we consider in this study targeted the search of new physics beyond the Standard Model in final states consisting a small number of jets in association with missing energy using $139~{\rm fb}^{-1}$ of data collected in the period of 2015-2018. In this analysis, events are required to have at least one jet with transverse momentum of $150$ GeV and no reconstructed isolated 'Loose' electrons or muons with $p_T > 7$ GeV and $|\eta| < 2.5$, tau leptons with $p_T > 10$ GeV and $|\eta| < 2.5$ or photons with $p_T > 10$ GeV and $|\eta| < 2.5$. The missing transverse  energy ($E_T^{\rm miss}$) is required to be larger than $200$ GeV. Twenty-six signal regions are defined depending on the cut on $E_T^{\rm miss}$: 13 inclusive signal region (IM0-IM12) and 13 exclusive signal regions (EM0-EM12).

\begin{figure}[!t]
    \centering
    \includegraphics[width=0.60\linewidth]{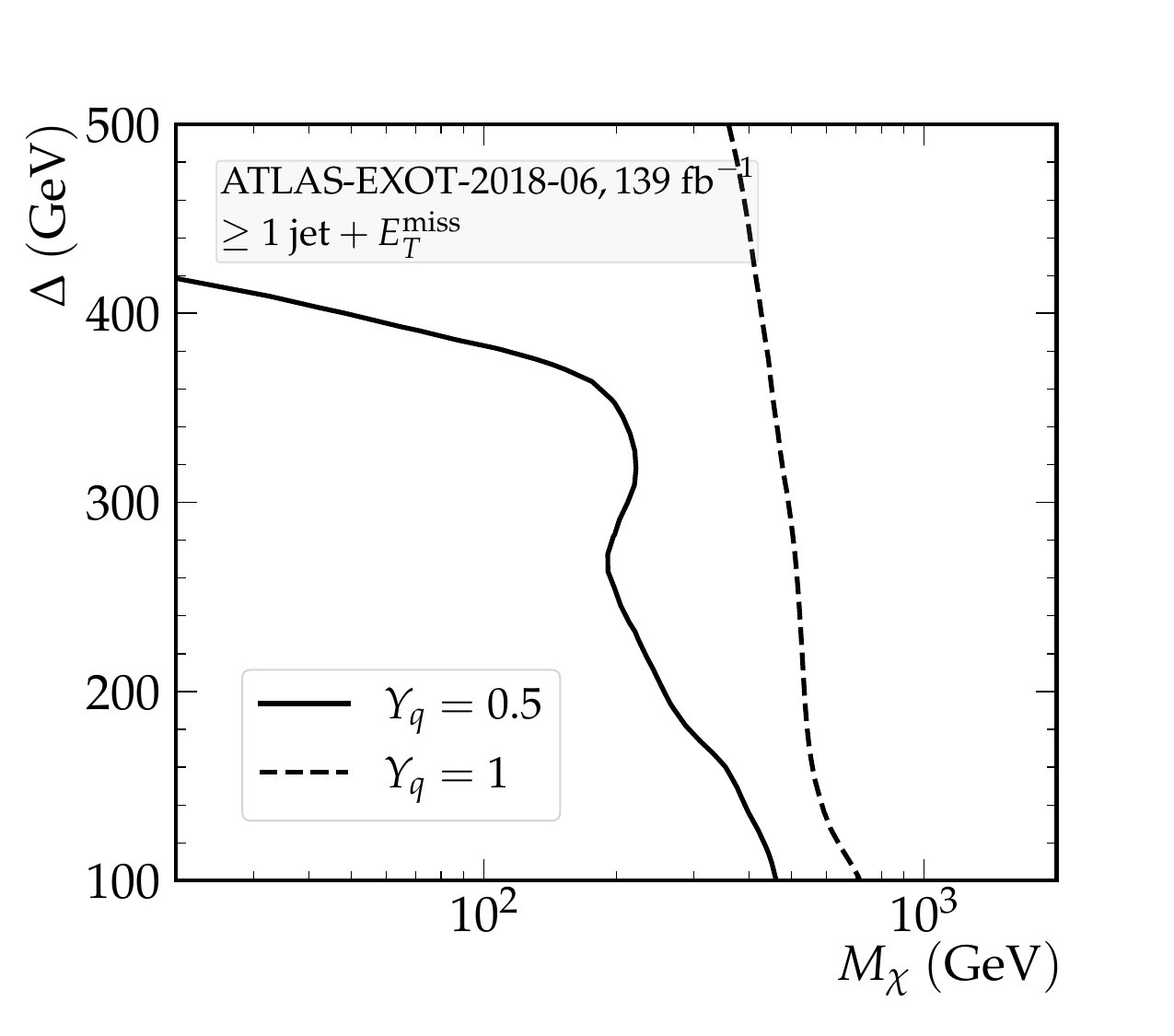}
\caption{The $95\%$ CL exclusion on the model projected on the plane of $(M_\chi, \Delta)$ for $Y_q=0.5$ (solid) and $Y_q$ (dashed). For the two cases, the contours correspond to ${\rm CL}_s^{95\%}$ which defines the exclusions at the $95\%$ confidence level. Here, we assumed that $Y_q=Y_u=Y_c$ for simplicity.}
\label{fig:LHC:exclusion}
\end{figure}

To estimate the bounds on the model parameter space arising from this search, we use a validated implementation, which is denoted by ATLAS-EXOT-2018-06, in the \textsc{MadAnalysis}~5 framework \cite{Conte:2012fm,Conte:2014zja,Dumont:2014tja,Conte:2018vmg}. The link to the analysis code along with the validation material can be found in ref. \cite{DVN/REPAMM_2023}. Theory predictions for the signal have been done using \textsc{MadGraph5$\_$aMC@NLO} version 3.5.0 \cite{Alwall:2014hca} with Leading Order (LO) matrix elements. The matrix elements have been convoluted with the \texttt{NNPDF31$\_$lo$\_$as$\_$0130} PDF set with $\alpha_s(M_Z)=0.130$ \cite{NNPDF:2017mvq} which can be found in \textsc{Lhapdf} version 6.4.0 \cite{Buckley:2014ana}. Note that our choice of $\alpha_s$ is adopted to capture some of the missing higher order corrections although the calculation is only accurate at LO. The generated events are then interfaced to \textsc{Pythia}~8 version 8309 \cite{Bierlich:2022pfr} to add parton showering and hadronization. All the jets were clustered with the anti-$k_t$ algorithm with a jet radius of $R=0.4$ \cite{Cacciari:2008gp} using \textsc{FastJet} version 3.40 \cite{Cacciari:2011ma}. Simplified detector modeling was achieved with the use of the Simplified Fast detector Simulation (SFS) module of \textsc{MadAnalysis}~5 \cite{Araz:2020lnp}. To estimate the exclusion bounds on the model we calculate the ${\rm CL}_{\rm s}$ using \texttt{Pyhf} \cite{pyhf_joss}. A point in the model parameter space is excluded at the $95\%$ if ${\rm CL}_{\rm s} > 0.95$. In fig.  \ref{fig:LHC:exclusion} we show the $95\%$ confidence level exclusions projected on the plane of $(M_\chi, \Delta)$ for two assumptions on the coupling: $Y_q = 0.5$ (solid) and $Y_q = 1$ (dashed). We can see that for $Y_q=1$ DM masses up to $800$ GeV are excluded with very small dependence on the mass splitting ($\Delta$). However, for the choice of $Y_q = 0.5$ the bounds get weaker given that total cross section behaves approximately as $Y_q^4$. Note that further improvements on the bounds can be made if one consider calculations at NLO (see ref. \cite{Arina:2020udz} for more details). We plan to improve our results in a future work \cite{Jueid:2024xyz}. 

\section{Benchmark scenarios for future collider analyses}
\label{sec:BPs}

\begin{table}[!t]
\setlength\tabcolsep{7pt}
\begin{center}
\begin{adjustbox}{max width=0.99\textwidth}\renewcommand{\arraystretch}{1.2}
\begin{tabular}{lccccc}
\multicolumn{1}{c} { Benchmark point } & Quantity & BP1 & BP2 & BP3 & BP4 \\
\toprule
\multirow{6}{*}{Parameters} & $Y_u$ & 0.4 & 0.4 & 0.0 & 0.4 \\
& $Y_c$ & 0.4 & 0.8 & 1.0 & 1.0 \\
& $Y_t$ & 0.4 & 1.2 & 2.0 & 0.8 \\
& $\lambda_3$ & 2.0 & 1.0 & 1.0 & 4.0 \\ 
& $M_\chi~({\rm GeV})$    & $500$  &   $200$ & $100$ & $600$ \\
& $\Delta~({\rm GeV})$    & $57$ & $650$ & $500$  & $250$  \\ 
& $\Lambda_{\rm pole}$ (TeV)  & 56 & 100 & 10 & 5.5 \\  [.3cm]

\multicolumn{6}{l}{ Branching ratios} \\
\toprule \\
\multirow{4}{*}{${\rm BR}(S\to q\chi)$} & ${\rm BR}(S \to u \chi)$ & $5.00 \times 10^{-1}$ & $7.60 \times 10^{-2}$ & $0.00 \times 10^{0}$ & $1.01 \times 10^{-1}$ \\
& ${\rm BR}(S \to c \chi)$ & $5.00 \times 10^{-1}$ & $3.03 \times 10^{-1}$ & $2.31 \times 10^{-1}$ & $6.32 \times 10^{-1}$ \\
& ${\rm BR}(S \to t \chi)$ & $0.00 \times 10^0$ & $6.21 \times 10^{-1}$ & $7.69 \times 10^{-1}$ & $2.67 \times 10^{-1}$ \\
& $\Gamma_{S}/M_{S}$ & $1.18 \times 10^{-4}$ & $3.64 \times 10^{-2}$ & $8.31 \times 10^{-2}$ & $7.92 \times 10^{-3}$  \\ [.3cm]
\multirow{4}{*}{${\rm BR}(t\to qX)$} & ${\rm BR}(t \to c H)$    & $1.02 \times 10^{-8}$ & $1.98 \times 10^{-8}$ & $3.69 \times 10^{-7}$ & $1.43 \times 10^{-7}$ \\
& ${\rm BR}(t \to u H)$    & $1.02 \times 10^{-8}$ & $4.95 \times 10^{-9}$ & $0.0$ & $2.29 \times 10^{-8}$ \\
& ${\rm BR}(t \to c Z)$    & $1.50 \times 10^{-8}$ & $1.79 \times 10^{-7}$ & $3.49 \times 10^{-6}$ & $5.92 \times 10^{-8}$ \\
& ${\rm BR}(t \to u Z)$    & $1.50 \times 10^{-8}$ & $4.48 \times 10^{-8}$ & $0.0$ & $9.48 \times 10^{-9}$ \\
\end{tabular}
\hspace{0.2cm}
\end{adjustbox}
\end{center}
\caption{Definition of the benchmark points. Here, we show the branching ratios ${\rm BR}(S\to \chi q)$ along with the width-to-mass ratio ($\Gamma_S/M_S$) of $S$. The FCNC branching ratios of the top quark, ${\rm BR}(t\to qH)$ and ${\rm BR}(t\to qZ)$, are also shown. For each benchmark point we display the energy scale, denoted by $\Lambda_{\rm pole}$, at which the perturbativity bound is violated at the one-loop order (more details about the renormalization group equation can be found in Appendix \ref{sec:RGEs}).}
\label{tab:BPs:definitions}
\end{table} 

\begin{table}[!t]
\setlength\tabcolsep{7pt}
\begin{center}
\begin{adjustbox}{max width=0.99\textwidth}\renewcommand{\arraystretch}{1.2}
\begin{tabular}{lccccc}
\multicolumn{1}{c} { Benchmark point } & Quantity & BP1 & BP2 & BP3 & BP4 \\ 
\toprule
\multirow{2}{*}{Dark matter} & $\Omega_{\chi} h^2$      & $1.18 \times 10^{-1}$ & $6.42 \times 10^{-2}$ & $8.58 \times 10^{-2}$ & $1.05 \times 10^{-1}$  \\
& $\sigma_{\rm SI}^p~({\rm pb})$ & $4.74 \times 10^{-11}$ & $3.51 \times 10^{-14}$ & $4.57 \times 10^{-13}$  & $2.97 \times 10^{-12}$  \\ [.3cm]
\multicolumn{6}{l}{Production cross sections [fb]} \\
\toprule
\multirow{12}{*}{$13.6~{\rm TeV}$} & $S\chi$ & $6.11 \times 10^1$ & $3.23 \times 10^1$ & $7.89 \times 10^1$ & $1.34 \times 10^1$ \\
& $S S^\dagger$ & $1.56 \times 10^{2}$ & $1.19 \times 10^1$ & $1.06 \times 10^2$ & $1.16 \times 10^1$ \\
& $S S + {\rm h.c.}$ & $1.79 \times 10^1$ & $1.45 \times 10^0$ & $4.80 \times 10^{-1}$ & $5.47 \times 10^0$ \\
& $\chi \chi H$ & $3.36 \times 10^{-4}$ & $2.65 \times 10^{-4}$ & $9.00 \times 10^{-4}$ & $4.94 \times 10^{-4}$ \\
& $\chi \chi Z$ & $1.82 \times 10^{-3}$ & $1.25 \times 10^{-2}$ & $1.48 \times 10^{-2}$ & $2.08 \times 10^{-3}$ \\
& $\chi S H$ & $5.35 \times 10^{-2}$ & $3.85 \times 10^{-3}$ & $1.11 \times 10^{-2}$ & $3.02 \times 10^{-2}$ \\
& $\chi S Z$ & $4.44 \times 10^{-2}$ & $2.27 \times 10^{-2}$ & $3.88 \times 10^{-2}$ & $1.12 \times 10^{-2}$ \\
& $S S^\dagger j$ & $2.19 \times 10^2$ & $1.64 \times 10^1$ & $1.46 \times 10^2$ & $1.63 \times 10^1$ \\
& $S S^\dagger \gamma$ & $1.02 \times 10^0$ & $1.10 \times 10^{-1}$ & $7.40 \times 10^{-1}$ & $1.10 \times 10^{-1}$ \\
& $S S^\dagger t$ & $8.21 \times 10^{-2}$ & $1.40 \times 10^{-1}$ & $1.01 \times 10^0$ & $4.50 \times 10^{-2}$ \\
& $S S^\dagger H$ & $4.80 \times 10^{-1}$ & $6.42 \times 10^{-3}$ & $7.69 \times 10^{-2}$ & $1.00 \times 10^{-1}$ \\
& $S S^\dagger Z$ & $2.40 \times 10^{-1}$ & $2.85 \times 10^{-2}$ & $1.80 \times 10^{-1}$ & $2.86 \times 10^{-2}$ \\
\toprule
\multirow{12}{*}{$100~{\rm TeV}$} & $S\chi$ & $3.41 \times 10^3$ & $2.32 \times 10^{3}$ & $6.53 \times 10^{3}$ & $1.57 \times  10^{3}$ \\
& $S S^\dagger$ & $28.82 \times 10^3$ & $4.63 \times 10^3$ & $21.36 \times 10^3$ & $4.61 \times 10^3$ \\
& $S S + {\rm h.c.}$ & $2.25 \times 10^2$ & $4.94 \times 10^1$ & $5.39 \times 10^1$ & $2.31 \times 10^2$ \\
& $\chi \chi H$ & $1.61 \times 10^{-2}$ & $1.01 \times 10^{-2}$ & $5.09 \times 10^{-2}$ & $4.69 \times 10^{-2}$ \\
& $\chi \chi Z$ & $9.91 \times 10^{-2}$ & $5.03 \times 10^{-1}$ &  $8.84 \times 10^{-1}$ & $2.04 \times 10^{-1}$ \\
& $\chi S H$ & $4.32 \times 10^0$ & $4.07 \times 10^{-1}$ & $1.39 \times 10^0$ & $5.06 \times 10^1$ \\
& $\chi S Z$ & $4.24 \times 10^0$ & $2.27 \times 10^0$ & $5.35 \times 10^0$ & $2.26 \times 10^0$ \\
& $S S^\dagger j$ & $58.65 \times 10^3$ & $10.36 \times 10^3$ & $43.92 \times 10^3$ & $10.32 \times 10^3$ \\
& $S S^\dagger \gamma$ & $1.38 \times 10^2$ & $2.48 \times 10^1$ & $8.91 \times 10^1$ & $2.75 \times 10^1$ \\
& $S S^\dagger t$ & $1.38 \times 10^1$ & $6.65 \times 10^1$ & $3.73 \times 10^2$ & $2.25 \times 10^1$ \\
& $S S^\dagger H$ & $1.28 \times 10^2$ & $3.64 \times 10^0$ & $2.24 \times 10^1$ & $5.84 \times 10^1$ \\
& $S S^\dagger Z$ & $2.65 \times 10^1$ & $6.66 \times 10^0$ & $2.16 \times 10^1$ & $6.70 \times 10^0$ \\
\end{tabular}
\hspace{0.2cm}
\end{adjustbox}
\end{center}
\caption{Some characteristics of the benchmark scenarios to be considered for future collider analyses.}
\label{tab:BPs}
\end{table} 

In this section, we present four benchmark points consistent with the current experimental bounds from LHC searches of new physics, top quark FCNC decays, Higgs boson couplings and cosmology. We also show the production rates for several processes which may be amenable to discovery at either the HL-LHC or the FCC-hh. The definition of the benchmark points is shown in Table \ref{tab:BPs:definitions} including the decay branching ratios of the colored scalar mediator and the top quark FCNC. Given the importance of the choice of $\lambda_3$ in our model, we also show the energy scale at which the perturbativity of the model is broken down (more details about the RGEs are shown in Appendix \ref{sec:RGEs}). Some of the  characteristics of the benchmark points are shown in Table \ref{tab:BPs}.  We give a few comments about the benchmark points: 

\paragraph{BP1.} This benchmark point is characterized by two main properties. First, all the Yukawa-type couplings ($Y_q$) are chosen to be equal, {\it i.e.} $Y_u = Y_c = Y_t = 0.4$. Second, we have chosen a small mass splitting between $\chi$ and $S$; $\Delta = 57$ GeV. For this choice of $M_\chi$, the dominant contribution to the relic abundance comes from the co-annihilation mechanism. We list the channels by their contribution to $(\Omega_\chi h^2)^{-1}$: $\chi S \to q g, q H, t$ contribute by about $65\%$, $\chi \chi \to q_\beta \bar{q}_\alpha$ contributes by about $21\%$ and $S S^\dagger \to W^+ W^-, gg$ contribute by about $7\%$. Given that the mass splitting is smaller than the top quark mass, the main decay channel of $S$ is into $u\chi$ and $c\chi$ with equal branching ratio of $50\%$ for each channel. The choice of $Y_u = Y_c$ leads to ${\rm BR}(t \to u X) = {\rm BR}(t \to c X)$ while the choice of $\lambda_3 = 2$ implies that ${\rm BR}(t\to q Z) \approx 1.5 \times {\rm BR}(t \to q H)$.  For this benchmark point, we find that the best channels to look for are the production of $\chi\chi$ in association with jets: $S(\to j\chi)\chi$, $S(\to j\chi) S^\dagger(\to j\chi)$, $S (\to j\chi) S^\dagger (\to j\chi) j$ and $S (\to j\chi) S^\dagger (\to j\chi) H (\to b\bar{b})$ where $j=u,\bar{u},c,\bar{c}$. There is also an important process to search for at hadron colliders which is the production of $SS^\dagger \gamma$ which leads to final states comprising of at least $2$ jets, a high-$p_T$ photon and large missing energy.

\paragraph{BP2.} Here we choose a normal hierarchy for the couplings, {\it i.e.} $Y_t > Y_c > Y_u$. Large mass splitting ($\Delta=650$ GeV) and a relatively light DM ($M_\chi = 200$ GeV) are chosen. Given this large mass splitting, the DM relic density is mainly due to the annihilation mechanism where annihilation into $t\bar{t}$ and $c\bar{t}+t\bar{c}$ contributes to the relic density by $55\%$ and $34\%$ respectively. In this benchmark scenario, the relic density of the $\chi$ particles form about $53\%$ of the total DM relic density in the universe. On the other hand, the top quark FCNC branching ratios satisfy ${\rm BR}(t\to cX) \approx 4 \times {\rm BR}(\to uX)$ and ${\rm BR}(t\to qZ) \approx 9.04 \times {\rm BR}(t\to qH)$. Due to this choice of couplings, the $S$ particle decays dominantly into $t\chi$ with BR of $62.1\%$ followed by $S\to c\chi$ with BR of $30.3\%$ while the decay $S\to u\chi$ is subleading with BR smaller than $10\%$. In this scenario, processes involving one top quark or more in association with large missing energy are the most prominent at hadron colliders. Here five processes may lead to interesting signatures: $S (\to t\chi) \chi$, $S (\to t\chi) S^\dagger(\to \bar{t}\chi)$, $S (\to t\chi) S (\to t\chi)+{\rm h.c.}$\footnote{Note that the production of $SS$ always dominates over $S^\dagger S^\dagger$ since its rate is initiated by valence $u$ quarks.}, $S (\to t\chi) S^\dagger (\to \bar{t}\chi) j$ and $S (\to t\chi) S^\dagger (\to \bar{t}\chi) t$. The latter channel is interesting as it leads to final states of $3$ top quarks and missing energy. There are other channels that involve the production of one or two hard jets initiated by $u$ and $c$ quarks in association with one or two top quarks. 

\paragraph{BP3.} In this benchmark point, we specifically choose the coupling to the $u$ quark to be exactly zero and assume the other couplings to be $Y_t = 2 \times Y_c = 2$ and $\lambda_3 = 1$. For the particle masses, we choose a DM of mass $100$ GeV and a mediator with mass of $600$ GeV. For this choice of mass and couplings, the most dominant decay of $S$ is into $t\chi$ with a branching ratio of $76.9\%$ followed by the decay into $c\chi$ with a branching ratio of $23.1\%$. The relic density of the $\chi$ particle in this benchmark point is about $71.5\%$ of the total DM relic density and is mainly due to the annihilation of $\chi$ into $c\bar{t}+t\bar{c}$ with a contribution of $92\%$. In this scenario, the branching ratios of $t\to uZ$ and $t\to uH$ are exactly zero while the decays involving charm quarks satisfy: ${\rm BR}(t\to cZ) \approx 9.46 \times {\rm BR}(t\to cH)$. In addition to processes like the production of one or two top quarks and jets in association with missing energy, this BP can be probed using processes involving Higgs bosons as well, {\it i.e.} $S (\to t \chi) S^\dagger (\to \bar{c}\chi) H$ and $\chi S(\to t\chi) H$. These two processes are advantageous since they have smaller associated backgrounds and can be used to connect top FCNCs and DM at hadron colliders.

\paragraph{BP4.} We select the couplings to satisfy $Y_c > Y_t > Y_u$ and $\lambda_3 = 4$. Moreover, we choose the DM to be quite heavy with a mass of $600$ GeV and a mass splitting of $250$ GeV. The dominant decay of $S$ is into $c\chi$ with a branching ratio of $62.3\%$ followed by $t\chi$ and $u\chi$ with branching ratios of $26.7\%$ and $10.1\%$ respectively. The top FCNC decays into $cX$ dominate over $uX$ with branching ratios satisfying ${\rm BR}(t\to c X) \approx 6.24 \times {\rm BR}(t\to uX)$ where the proportionality factor is approximately equal to $(Y_c/Y_u)^2$. In this BP, the branching ratio of top FCNC decay into $H+q$ is larger than into $Z+q$ and satisfy ${\rm BR}(t\to qH) \approx 2.41 \times {\rm BR}(t\to qZ)$. The relic density in this BP is mainly due to the annihilation of $\chi$ into $c\bar{t}$, $t\bar{c}$, $t\bar{t}$ and $c\bar{c}$ with a combined contribution of about $86\%$. This choice leads to $\Omega_\chi h^2/\Omega_{\rm Planck} h^2 \approx 87.1\%$. The heavy DM scenario in this BP leads to smaller cross sections for the production of $\chi$ and $S$ in hadron colliders as compared to the rates for the other BPs. 

\section{Conclusions}
\label{sec:conclusions}

In this work, we have suggested a minimal simplified model that simultaneously addresses the DM problem and generates {\it nonzero} rates for top quark FCNC decays. In this model, the SM is extended by two $SU(2)_L$ singlets: a colored scalar mediator ($S$) that carries the same quantum numbers as a right-handed up-type quark, and a Majorana fermion ($\chi$) which plays the role of the DM candidate. The two extra states are odd under an ad-hoc $Z_2$ symmetry which is imposed to guarantee the stability of $\chi$ provided that $M_\chi \leq M_S$. Since the colored scalar mediator couples to all the quark generations, {\it nonzero} rates for top quark FCNC decays are induced through the loops of $S$ and $\chi$.  Using examples of two interesting top quark FCNC decays, {\it i.e.} $t\to qH$ and $t\to qZ$, we have comprehensively analyzed the contribution of these two extra states on the corresponding branching ratios. First, we find that the top quark FCNC branching ratios do not depend on the light quark masses but only on their coupling to $\chi$ and $S$ (denoted by $Y_q$). Second, we find that the branching ratios ${\rm BR}(t\to qH)$ and ${\rm BR}(t\to q Z)$ are related by phase space factors and the coupling of $S$ to the SM Higgs boson doublet. We then analyzed the DM relic density in this model which is mainly due to the annihilation of the $\chi$ particles into quarks unless the mass splitting between $\chi$ and $S$ is small in which case co-annihilation into SM particles starts to dominate. The bounds from the LHC searches of DM in monojet are analyzed. After analyzing all these constraints, we have defined four benchmark points that can lead to high discovery potential at the HL-LHC and FCC-hh. We discussed in details the characteristics of these benchmark points and the different methods to probe them at high-energy colliders. This work provides a novel interesting connection between top quark FCNC, the DM problem and collider searches of new physics BSM.

\section*{Acknowledgments}
The work of A.J. is supported by the Institute for Basic Science (IBS) under the project code, IBS-R018-D1. The work of S. K. was supported by the JSPS KAKENHI Grant No. 20H00160 and No. 23K17691. A.J. would like to thank the CERN Theory Department for its hospitality where part of this work has been done.

\appendix

\section{Impact on Higgs boson couplings}
\label{sec:Higgs}

\begin{figure}[!t]
\centering
\includegraphics[width=0.49\linewidth]{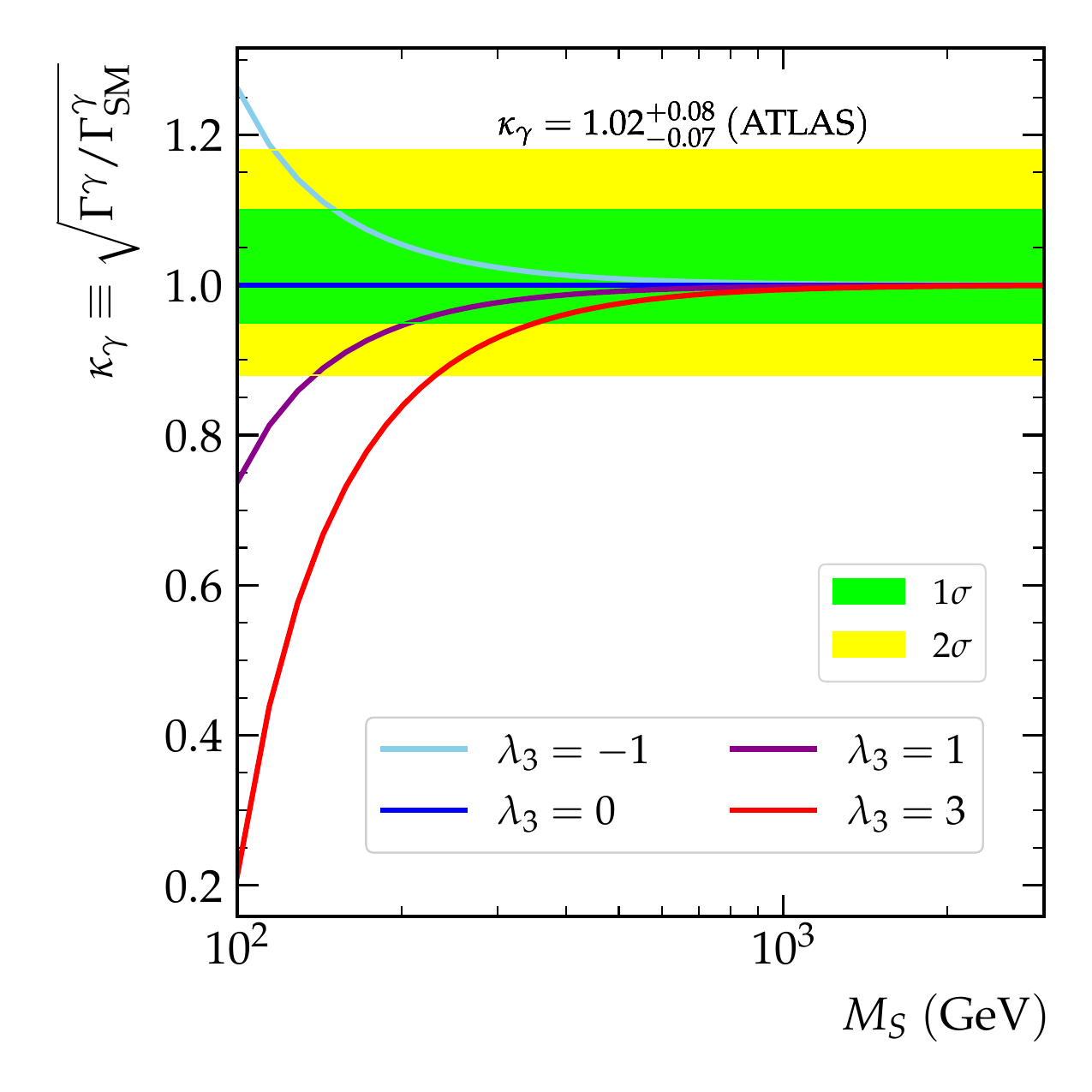}
\hfill
\includegraphics[width=0.49\linewidth]{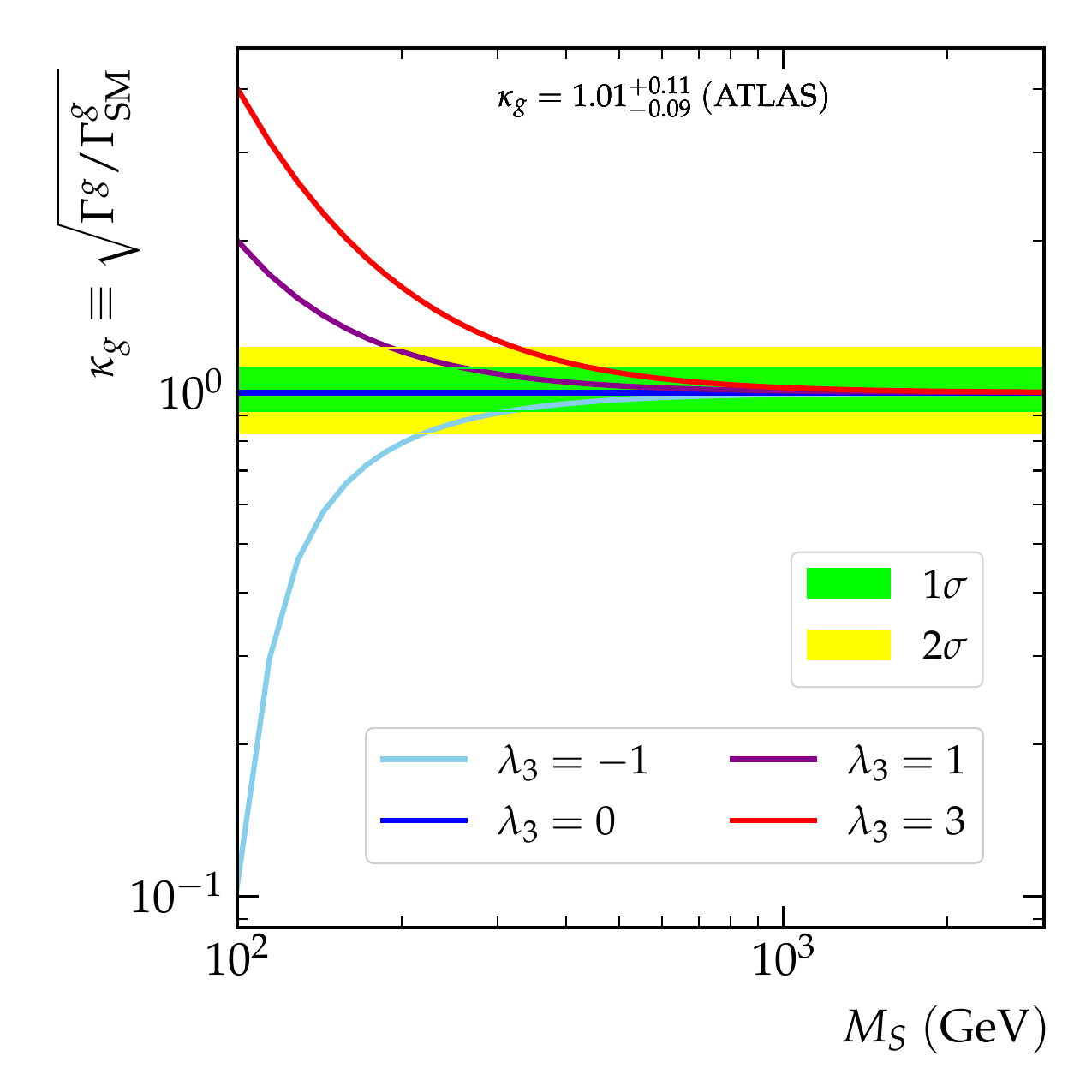}
\caption{The dependence of $\kappa_\gamma$ (left) and $\kappa_g$ (right) on the colored scalar mass ($M_S$) for $\lambda_3 = -1$ (turquoise), $\lambda_3 = 0$ (blue), $\lambda_3 = 1$ (purple) and $\lambda_3 = 3$ (red). In the same plots we show the best-fit values of $\kappa_\gamma = 1.02^{+0.08}_{-0.07}$ and $\kappa_g = 1.01^{+0.11}_{-0.09}$ along with the $1\sigma$ (green) and $2\sigma$ (yellow) bands as was reported on by the ATLAS collaboration \cite{ATLAS:2022tnm}.}
\label{fig:kappagg}
\end{figure}

As pointed out in section \ref{sec:theory}, the Higgs boson couplings get contributions from the parameters of this model. Here, we show explicitly the impact of the Higgs measurements on the allowed range of the model's parameters. There are two different categories of decay channels that can be affected by this model: bosonic decays into $\gamma\gamma$ and $gg$ and fermionic decays into $u\bar{u}, c\bar{c}$. In this appendix, we do not consider the contribution to bosonic decay processes like $ZZ^*, WW^*$ since it was found that they are small and do not go beyond a few percents, see refs. \cite{Arhrib:2015hoa,Kanemura:2016sos,Kanemura:2017wtm}. The main aim of this section is to evaluate the following ratio

\begin{eqnarray}
\kappa_i = \sqrt{\frac{\Gamma(H \to i)}{\Gamma(H\to i)^{\rm SM}}},
\label{eq:kappa}
\end{eqnarray}
where $i=g,\gamma,u,c$. We start with the bosonic decay channels. In this model, the new contribution to these decay widths depends solely on the mass of the colored mediator ($M_S$) and the quartic coupling ($\lambda_3$). The partial decay widths for $\gamma\gamma$ and $gg$ channels are given by
\begin{eqnarray}
\Gamma(H\to \gamma\gamma) &=& \frac{G_F \alpha_{\rm EM}^2 m_H^3}{128 \sqrt{2} \pi^3} \bigg|\sum_f Q_f^2 N_{cf} A_{1/2}(\tau_f) + A_{1}(\tau_W) + N_{cS} Q_S^2 \frac{\lambda_3 v^2}{2 M_S^2} A_{0}(\tau_S) \bigg|^2, \nonumber \\
\Gamma(H\to gg) &=& \frac{G_F \alpha_s^2 m_H^3}{64 \sqrt{2} \pi^3} \bigg|\sum_f A_{1/2}(\tau_f) + \frac{\lambda_3 v^2}{2 M_S^2} A_{0}(\tau_S) \bigg|^2,
\end{eqnarray}
with $N_{cX}=3$ being the number of colors for the quarks and the scalar mediator, $\tau_i = m_H^2/(4 m_i^2)$, $A_{1/2}(\tau) = 2 \tau^{-2} (\tau + (\tau - 1) f(\tau))$, $A_{1}(\tau) = -\tau^{-2} (2 \tau^2 + 3 \tau + 3 (2 \tau - 1) f(\tau))$, $A_{0}(\tau) = - \tau^{-2} (\tau - f(\tau))$, and $f(\tau)$ is the one-loop function which can be found in e.g. ref. \cite{Djouadi:2005gi}. 
In the SM, the contribution of the $W$-boson to $\Gamma(H\to \gamma\gamma)$ is dominant as compared to the contribution of the top quark and it comes with an opposite sign.  The contribution of the colored scalar is mainly controlled by the value of $\lambda_3$. We can see that there are destructive (constructive) interference for positive (negative) values of $\lambda_3$ with the dominant $W$-boson contribution. The situation is different for the case of $H \to gg$ since the only dominant contribution in the SM is that of the top quark. The new scalar contribution comes with the same sign as the top quark contribution for $\lambda_3 > 0$ leading to enhancement while it reduces the rate of $H \to  gg$ for negative $\lambda_3$. These features can be clearly seen in fig.  \ref{fig:kappagg} where we show the dependence of $\kappa_\gamma$ (left) and $\kappa_g$ on $M_S$ for different values of $\lambda_3$. We can see that $\kappa_g$ and $\kappa_\gamma$ are anticorrelated in this model since for example the new scalar loops induce positive~(negative) contributions to $\kappa_\gamma$~($\kappa_g$) when $\lambda_3 < 0$\footnote{The choice of a negative value of $\lambda_3$ may lead to vacuum configurations that break the color symmetry. Here we only show the results for comparison since an analysis of the color-breaking minima is beyond the scope of this work.}. To compare with the experimental data, we also show the recent measurements of $\kappa_\gamma$ and $\kappa_g$ reported on by the ATLAS collaboration \cite{ATLAS:2022tnm}. We can see that the recent measurements of $\kappa_\gamma$ and $\kappa_g$ do not prefer light scalars as masses of order $200$--$300$ GeV are excluded for all but $\lambda_3 \approx 0$. 

We turn now into a brief discussion of the contribution of the new states to the fermionic rates {\it i.e.,} $H\to u\bar{u}$ and $H \to c\bar{c}$. The partial width for these channels is given by
\begin{eqnarray}
\Gamma(H \to q\bar{q}) \equiv \Gamma(H \to q\bar{q})_{\rm N3LO} + \Delta \Gamma(H \to q\bar{q})_{\rm NP},
\end{eqnarray}
where $\Gamma(H \to q\bar{q})_{\rm N3LO}$ is the decay width in the SM calculated at N3LO including renormalized running quark masses \cite{Chetyrkin:1996sr,Chetyrkin:1997vj} and $\Delta \Gamma(H \to q\bar{q})_{\rm NP}$ is the model contribution to the decay width which is given by 
\begin{eqnarray}
\Delta \Gamma(H \to q\bar{q})_{\rm NP} = \frac{6 m_H m_q}{16 \pi v} \bigg[{\rm Re}(f_L + \delta f_L) + {\rm Re}(f_R + \delta f_R)\bigg], 
\end{eqnarray}
with $f_{L,R}$ being the one-loop form factors which depend on $Y_q$, $\lambda_3$, $M_\chi$ and $M_S$ and they are given by

\begin{eqnarray}
f_L = \frac{3 \lambda_3 m_q v Y_q^2}{16 \pi^2} C_2(m_q^2, m_H^2, m_q^2, M_\chi^2, M_S^2, M_S^2), \nonumber \\
f_R = \frac{3 \lambda_3 m_q v Y_q^2}{16 \pi^2} C_1(m_q^2, m_H^2, m_q^2, M_\chi^2, M_S^2, M_S^2).
\end{eqnarray}

\begin{figure}[!t]
\centering
\includegraphics[width=0.32\linewidth]{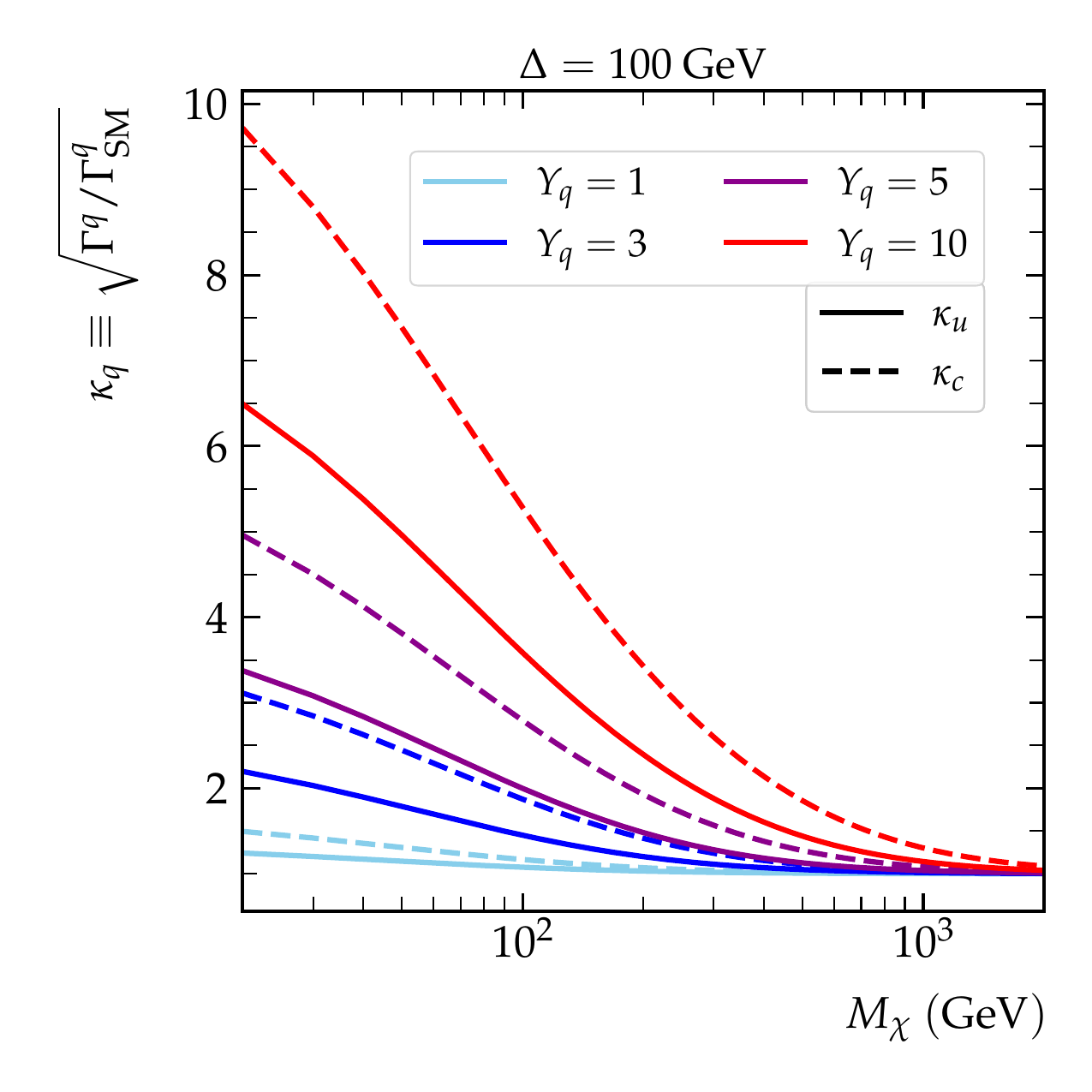}
\hfill
\includegraphics[width=0.32\linewidth]{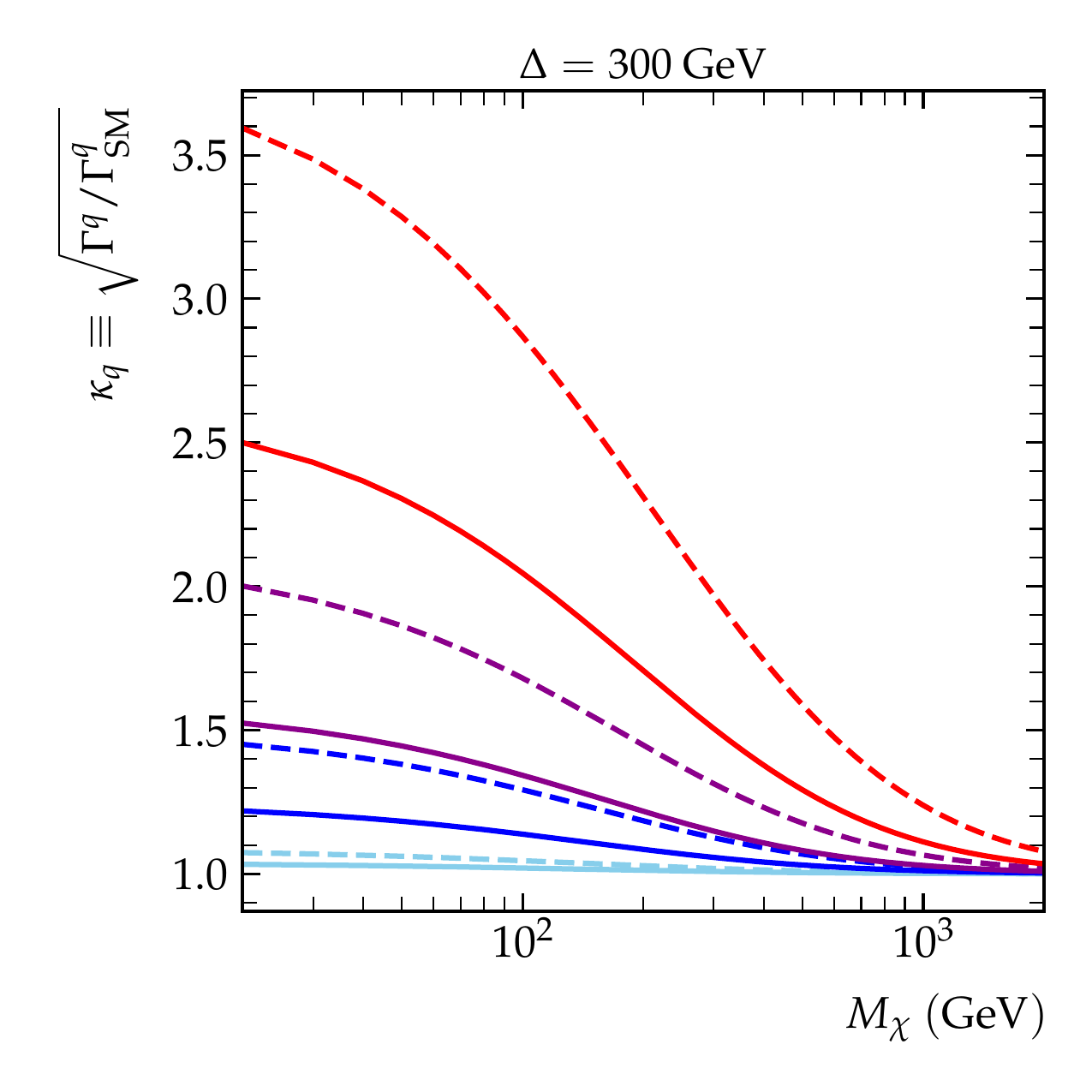}
\hfill
\includegraphics[width=0.32\linewidth]{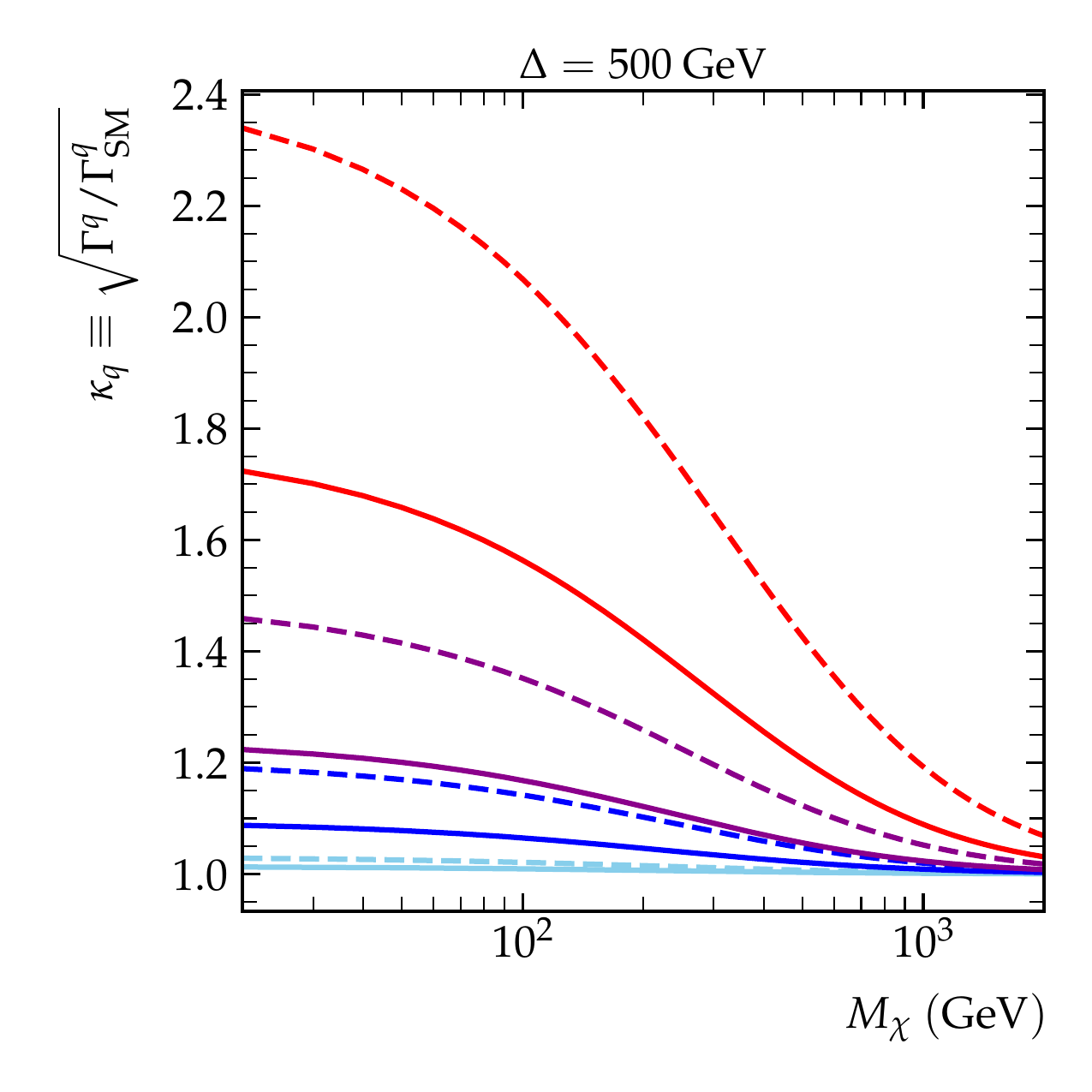}
\caption{Dependence of $\kappa_q$ on the dark-matter mass ($M_\chi$) for $\Delta = 100~{\rm GeV}$ (left panel), $\Delta = 300~{\rm GeV}$ (middle panel) and $\Delta = 500~{\rm GeV}$ (right panel). Here, we show the results for $\kappa_c$ (solid lines) and $\kappa_u$ (dashed lines). For each panel, the results are shown for $Y_q = 1$ (purple), $Y_q = 3$ (blue), $Y_q = 5$ (purple) and $Y_q = 10$ (red).}
\label{fig:kappaff}
\end{figure}

$\delta f_L$ and $\delta f_R$ are the counterterms given by \cite{Denner:1991kt}
\begin{eqnarray}
    \delta f_L &=& \frac{m_q e }{2 s_W M_W} \bigg[1 + \delta Z_e - \frac{\delta s_W}{s_W} + \frac{\delta m_q}{m_q} - \frac{\delta M_W}{M_W} + \frac{1}{2} \delta Z_H + \frac{1}{2} \bigg(\delta Z_{qq}^R + \delta Z_{qq}^{L,\dagger} \bigg) \bigg], \nonumber \\
    \delta f_R &=& \frac{m_q e }{2 s_W M_W} \bigg[1 + \delta Z_e - \frac{\delta s_W}{s_W} + \frac{\delta m_q}{m_q} - \frac{\delta M_W}{M_W} + \frac{1}{2} \delta Z_H + \frac{1}{2} \bigg(\delta Z_{qq}^L + \delta Z_{qq}^{R,\dagger} \bigg) \bigg], 
\end{eqnarray}
the renormalization constants $\delta Z$ are easily calculated:
\begin{eqnarray}
    \delta M_Z^2 &=& \frac{- \alpha s_W^2}{c_W^2 \pi} \bigg(\frac{-2}{3} A_0(M_S^2) + \frac{4}{3} B_{00}(M_Z^2, M_S^2, M_S^2) \bigg), \quad \delta M_W^2 = 0, \nonumber  \\
    \delta Z_e &=& \frac{1}{2} \bigg(-\delta Z_{\gamma\gamma} - \frac{s_W}{c_W} \delta Z_{Z\gamma} \bigg), \quad \delta s_W = \frac{c_W^2}{2 s_W} \bigg(-\frac{\delta M_W^2}{M_W^2} + \frac{\delta M_Z^2}{M_Z^2}\bigg), \nonumber \\
    \delta Z_{\gamma\gamma} &=& \frac{4\alpha}{3 \pi} \frac{\partial}{\partial q^2} B_{00}(q^2, M_S^2, M_S^2)\bigg|_{q^2=0}, \quad \delta Z_H = \frac{3 \alpha \lambda_3^2 M_W^2}{\pi s_W^2} \frac{\partial}{\partial q^2} B_{0}(q^2, M_S^2, M_S^2)\bigg|_{q^2=m_H^2}, \nonumber \\ 
    \delta Z_{Z\gamma} &=& \frac{- 2 \alpha s_W}{c_W M_Z^2 \pi} \bigg(\frac{-2}{3} A_0(M_S^2) + \frac{4}{3} B_{00}(M_Z^2, M_S^2, M_S^2) \bigg), \\
    \delta Z_{qq}^L &=& \frac{m_q^2 Y_q^2}{16 \pi^2} \frac{\partial}{\partial q^2} B_{1}(q^2, M_\chi^2, M_S^2)\bigg|_{q^2=m_q^2}, \quad \delta m_q = \frac{-1}{32} \frac{m_q Y_q^2}{\pi^2} B_1(m_q^2, M_\chi^2, M_S^2), \nonumber \\
    \delta Z_{qq}^R &=& \frac{Y_q^2}{16 \pi^2} \bigg(B_1(m_q^2, M_\chi^2, M_S^2) + m_q^2 \frac{\partial}{\partial q^2} B_1(q^2, M_\chi^2, M_S^2)\bigg|_{q^2=m_q^2} \bigg). \nonumber
\end{eqnarray}
The numerical evaluation of the fermionic decay widths has been done using a customised \texttt{Python} code employing \textsc{PyCollier} \cite{Bahl:2023eau}\footnote{\textsc{PyCollier} is a \texttt{Python} wrapper of the \textsc{Collier} library \cite{Denner:2016kdg}.}. The results of our calculations are shown in fig.  \ref{fig:kappaff} where we show $\kappa_u$ (solid) and $\kappa_c$ (dashed) as a function of $M_\chi$ for $\Delta = 100$ GeV (left), $\Delta=300$ GeV (middle) and $\Delta=500$ GeV (right). We can see that unless the couplings $Y_q$ are very large, {\it i.e.} $Y_q > 5$ or so, the corrections to $\kappa_q$ are always small. For large $M_\chi$, all the corrections are decoupling and $\kappa_q$ reach their SM values.

\section{Contribution to the $\rho$ parameter}
\label{sec:rho}

\begin{figure}[!t]
    \centering
    \includegraphics[width=0.6\linewidth]{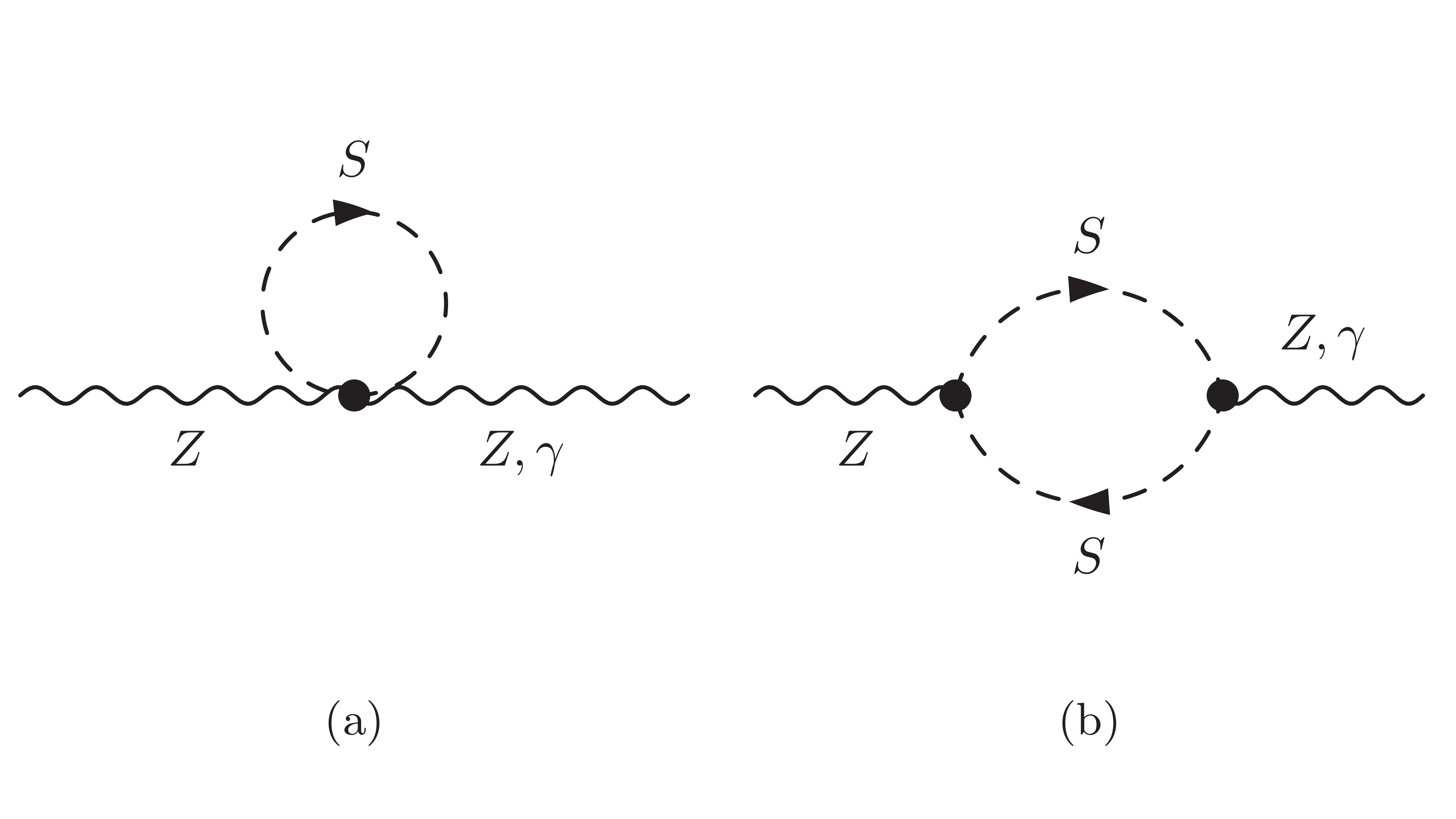}
    \vspace{-0.8cm}
    \caption{Examples of Feynman diagrams that contribute to $ZZ$ and $Z\gamma$ self energies at the one-loop order.}
    \label{fig:FD:rho}
\end{figure}

In this section, we demonstrate that our model give a zero contribution to the $\rho$ parameter (this was mentioned in section \ref{sec:theory}). We explicitly calculate the contribution to the $\rho$ parameter at the one-loop order where the leading order Feynman diagrams in this model are shown in fig.  \ref{fig:FD:rho}. The expression of $\Delta \rho$ is given by
\begin{eqnarray}
    \Delta \rho = \bigg(\frac{\Pi_{ZZ}(0)}{M_Z^2} - \frac{\Pi_{WW}(0)}{M_W^2} - \frac{2 s_W}{c_W} \frac{\Pi_{Z\gamma}(0)}{M_Z^2} \bigg),
\end{eqnarray}
where $\Pi_{VV'}(q^2)$ is the contribution to the 1PI two-point function and $s_W = \sin^2\theta_W$ is the sine of the Weinberg mixing angle.
Since the colored scalar mediator is a singlet under $SU(2)_L$, its contribution to the $W$--boson self energy is exactly zero. Using \textsc{FeynArts} and \textsc{FormCalc}, we have:
\begin{eqnarray}
    \Pi_{ZZ}(0) &=& -\frac{g_1^2 s_W^2}{\pi^2} \bigg(\frac{1}{6} A_0(M_S^2) - \frac{1}{3} B_{00}(0, M_S^2, M_S^2) \bigg), \nonumber \\
    \Pi_{Z\gamma}(0) &=& \frac{g_1^2 s_W c_W}{\pi^2} \bigg(\frac{1}{6} A_0(M_S^2) - \frac{1}{3} B_{00}(0, M_S^2, M_S^2) \bigg), 
\end{eqnarray}
where $A_0(x)$ and $B_{00}(0, x, x)$ are the one- and two-point scalar loop functions and $g_1 = e / c_W$ is the $U(1)_Y$ gauge coupling. We get 
\begin{eqnarray}
    \Delta\rho = -\frac{3 g_1^2 s_W^2}{\pi^2 M_Z^2} \bigg(\frac{1}{6} A_0(M_S^2) - \frac{1}{3} B_{00}(0, M_S^2, M_S^2) \bigg).
\end{eqnarray}
We must note that $\Delta\rho$ given above is free of UV divergences since the UV divergent part of the two Passarino-Veltman functions is 
\begin{eqnarray}
    {\rm Div}[A_{0}(m^2)] \equiv m^2 \Delta, \qquad {\rm Div}[B_{00}(p^2, m_1^2, m_2^2)] \equiv \bigg(\frac{m_1^2+m_2^2}{4} - \frac{p^2}{2}\bigg) \Delta,
\end{eqnarray}
where $\Delta = 2/\epsilon + \log(4\pi) - \gamma_E$. In our case the UV divergent part of $A_0$ and $B_{00}$ satisfy ${\rm Div}[A_0] = 2 \times {\rm Div}[B_{00}]$ which imply that ${\rm Div}[\Delta \rho]=0$.

\section{Renormalization group equations and high-energy behavior}
\label{sec:RGEs}

In this section we show the details of the renormalization group equations (RGEs) relevant for the analysis of section \ref{sec:BPs}. The beta function for a parameter $X$ is given by 
\begin{eqnarray}
    \beta\left(X\right) \equiv \mu \frac{d X}{d \mu}\equiv \frac{1}{\left(4 \pi\right)^{2}}\beta^{(1)}(X)+ \frac{1}{\left(4 \pi\right)^{4}}\beta^{(2)}(X) + \cdots,
\end{eqnarray}
where $\beta^{(1)}(X)$ and $\beta^{(2)}$ refer to the beta functions at the one-loop and two-order respectively. Higher order corrections are encoded in the $\cdots$. The calculation of the beta functions was performed using \texttt{PyR@Te} version 3.0 \cite{Sartore:2020gou}. Below we give the expression of the beta functions for the new parameters of the model at the one- and the two-loop orders. In what follows, $L_t = (Y_u, Y_c, Y_t)^T$. 

\subsection{Dark-matter couplings}
{\allowdisplaybreaks
\begin{align*}
\begin{autobreak}
\beta^{(1)}(L_t) =

+ Y_u^{\dagger} Y_u L_t

+ 4 L_t L_t^{\dagger} L_t

+ 2 \tr\left(L_t^{\dagger} L_t \right) L_t

-  \frac{4}{5} g_1^{2} L_t

- 4 g_3^{2} L_t,
\end{autobreak}
\end{align*}
\begin{align*}
\begin{autobreak}
\beta^{(2)}(L_t) =

-  \frac{1}{4} Y_u^{\dagger} Y_u Y_u^{\dagger} Y_u L_t

-  \frac{1}{4} Y_u^{\dagger} Y_d Y_d^{\dagger} Y_u L_t

-  \frac{3}{2} L_t L_t^{\dagger} Y_u^{\dagger} Y_u L_t

+ 5 L_t L_t^{\dagger} L_t L_t^{\dagger} L_t

-  \frac{9}{2} \tr\left(Y_u^{\dagger} Y_u \right) Y_u^{\dagger} Y_u L_t

- 3 \tr\left(Y_u^{\dagger} Y_u L_t L_t^{\dagger} \right) L_t

-  \frac{9}{2} \tr\left(Y_d^{\dagger} Y_d \right) Y_u^{\dagger} Y_u L_t

-  \frac{3}{2} \tr\left(Y_e^{\dagger} Y_e \right) Y_u^{\dagger} Y_u L_t

- 12 \tr\left(L_t^{\dagger} L_t L_t^{\dagger} L_t \right) L_t

- 12 \tr\left(L_t^{\dagger} L_t \right) L_t L_t^{\dagger} L_t

- 8 \lambda_3 Y_u^{\dagger} Y_u L_t

- 128 \lambda_2 L_t L_t^{\dagger} L_t

+ 128 \lambda_2^{2} L_t

+ 4 \lambda_3^{2} L_t

+ \frac{49}{120} g_1^{2} Y_u^{\dagger} Y_u L_t

+ \frac{51}{8} g_2^{2} Y_u^{\dagger} Y_u L_t

-  \frac{16}{3} g_3^{2} Y_u^{\dagger} Y_u L_t

+ \frac{268}{15} g_1^{2} L_t L_t^{\dagger} L_t

+ \frac{124}{3} g_3^{2} L_t L_t^{\dagger} L_t

+ \frac{4}{3} g_1^{2} \tr\left(L_t^{\dagger} L_t \right) L_t

+ \frac{20}{3} g_3^{2} \tr\left(L_t^{\dagger} L_t \right) L_t

+ \frac{232}{225} g_1^{4} L_t

-  \frac{64}{15} g_1^{2} g_3^{2} L_t

-  \frac{181}{9} g_3^{4} L_t.
\end{autobreak}
\end{align*}
}

\subsection{Quartic couplings}
{\allowdisplaybreaks

\begin{align*}
\begin{autobreak}
\beta^{(1)}(\lambda_1) =

+ 24 \lambda_1^{2}

+ 12 \lambda_3^{2}

-  \frac{9}{5} g_1^{2} \lambda_1

- 9 g_2^{2} \lambda_1

+ \frac{27}{200} g_1^{4}

+ \frac{9}{20} g_1^{2} g_2^{2}

+ \frac{9}{8} g_2^{4}

+ 12 \lambda_1 \tr\left(Y_u^{\dagger} Y_u \right)

+ 12 \lambda_1 \tr\left(Y_d^{\dagger} Y_d \right)

+ 4 \lambda_1 \tr\left(Y_e^{\dagger} Y_e \right)

- 6 \tr\left(Y_u^{\dagger} Y_u Y_u^{\dagger} Y_u \right)

- 6 \tr\left(Y_d^{\dagger} Y_d Y_d^{\dagger} Y_d \right)

- 2 \tr\left(Y_e^{\dagger} Y_e Y_e^{\dagger} Y_e \right),
\end{autobreak}
\end{align*}
\begin{align*}
\begin{autobreak}
\beta^{(2)}(\lambda_1) =

- 312 \lambda_1^{3}

- 120 \lambda_1 \lambda_3^{2}

- 96 \lambda_3^{3}

+ \frac{108}{5} g_1^{2} \lambda_1^{2}

+ 108 g_2^{2} \lambda_1^{2}

+ \frac{128}{5} g_1^{2} \lambda_3^{2}

+ 128 g_3^{2} \lambda_3^{2}

+ \frac{2063}{200} g_1^{4} \lambda

+ \frac{117}{20} g_1^{2} g_2^{2} \lambda

-  \frac{73}{8} g_2^{4} \lambda

+ \frac{24}{5} g_1^{4} \lambda_3

-  \frac{3747}{2000} g_1^{6}

-  \frac{1789}{400} g_1^{4} g_2^{2}

-  \frac{289}{80} g_1^{2} g_2^{4}

+ \frac{305}{16} g_2^{6}

- 144 \lambda_1^{2} \tr\left(Y_u^{\dagger} Y_u \right)

- 144 \lambda_1^{2} \tr\left(Y_d^{\dagger} Y_d \right)

- 48 \lambda_1^{2} \tr\left(Y_e^{\dagger} Y_e \right)

- 48 \lambda_3^{2} \tr\left(L_t^{\dagger} L_t \right)

+ \frac{17}{2} g_1^{2} \lambda_1 \tr\left(Y_u^{\dagger} Y_u \right)

+ \frac{5}{2} g_1^{2} \lambda_1 \tr\left(Y_d^{\dagger} Y_d \right)

+ \frac{15}{2} g_1^{2} \lambda_1 \tr\left(Y_e^{\dagger} Y_e \right)

+ \frac{45}{2} g_2^{2} \lambda_1 \tr\left(Y_u^{\dagger} Y_u \right)

+ \frac{45}{2} g_2^{2} \lambda_1 \tr\left(Y_d^{\dagger} Y_d \right)

+ \frac{15}{2} g_2^{2} \lambda_1 \tr\left(Y_e^{\dagger} Y_e \right)

+ 80 g_3^{2} \lambda_1 \tr\left(Y_u^{\dagger} Y_u \right)

+ 80 g_3^{2} \lambda_1 \tr\left(Y_d^{\dagger} Y_d \right)

-  \frac{171}{100} g_1^{4} \tr\left(Y_u^{\dagger} Y_u \right)

+ \frac{9}{20} g_1^{4} \tr\left(Y_d^{\dagger} Y_d \right)

-  \frac{9}{4} g_1^{4} \tr\left(Y_e^{\dagger} Y_e \right)

+ \frac{63}{10} g_1^{2} g_2^{2} \tr\left(Y_u^{\dagger} Y_u \right)

+ \frac{27}{10} g_1^{2} g_2^{2} \tr\left(Y_d^{\dagger} Y_d \right)

+ \frac{33}{10} g_1^{2} g_2^{2} \tr\left(Y_e^{\dagger} Y_e \right)

-  \frac{9}{4} g_2^{4} \tr\left(Y_u^{\dagger} Y_u \right)

-  \frac{9}{4} g_2^{4} \tr\left(Y_d^{\dagger} Y_d \right)

-  \frac{3}{4} g_2^{4} \tr\left(Y_e^{\dagger} Y_e \right)

- 3 \lambda_1 \tr\left(Y_u^{\dagger} Y_u Y_u^{\dagger} Y_u \right)

- 18 \lambda_1 \tr\left(Y_u^{\dagger} Y_u L_t L_t^{\dagger} \right)

- 42 \lambda_1 \tr\left(Y_u^{\dagger} Y_d Y_d^{\dagger} Y_u \right)

- 3 \lambda_1 \tr\left(Y_d^{\dagger} Y_d Y_d^{\dagger} Y_d \right)

-  \lambda_1 \tr\left(Y_e^{\dagger} Y_e Y_e^{\dagger} Y_e \right)

-  \frac{8}{5} g_1^{2} \tr\left(Y_u^{\dagger} Y_u Y_u^{\dagger} Y_u \right)

+ \frac{4}{5} g_1^{2} \tr\left(Y_d^{\dagger} Y_d Y_d^{\dagger} Y_d \right)

-  \frac{12}{5} g_1^{2} \tr\left(Y_e^{\dagger} Y_e Y_e^{\dagger} Y_e \right)

- 32 g_3^{2} \tr\left(Y_u^{\dagger} Y_u Y_u^{\dagger} Y_u \right)

- 32 g_3^{2} \tr\left(Y_d^{\dagger} Y_d Y_d^{\dagger} Y_d \right)

+ 30 \tr\left(Y_u^{\dagger} Y_u Y_u^{\dagger} Y_u Y_u^{\dagger} Y_u \right)

+ 12 \tr\left(Y_u^{\dagger} Y_u Y_u^{\dagger} Y_u L_t L_t^{\dagger} \right)

- 6 \tr\left(Y_u^{\dagger} Y_u Y_u^{\dagger} Y_d Y_d^{\dagger} Y_u \right)

- 6 \tr\left(Y_u^{\dagger} Y_d Y_d^{\dagger} Y_d Y_d^{\dagger} Y_u \right)

+ 30 \tr\left(Y_d^{\dagger} Y_d Y_d^{\dagger} Y_d Y_d^{\dagger} Y_d \right)

+ 10 \tr\left(Y_e^{\dagger} Y_e Y_e^{\dagger} Y_e Y_e^{\dagger} Y_e \right),
\end{autobreak}
\end{align*}
\begin{align*}
\begin{autobreak}
\beta^{(1)}(\lambda_2) =

+ 112 \lambda_2^{2}

+ 2 \lambda_3^{2}

-  \frac{16}{5} g_1^{2} \lambda_2

- 16 g_3^{2} \lambda_2

+ \frac{8}{75} g_1^{4}

+ \frac{4}{15} g_1^{2} g_3^{2}

+ \frac{13}{24} g_3^{4}

+ 8 \lambda_2 \tr\left(L_t^{\dagger} L_t \right)

- 2 \tr\left(L_t^{\dagger} L_t L_t^{\dagger} L_t \right),
\end{autobreak}
\end{align*}
\begin{align*}
\begin{autobreak}
\beta^{(2)}(\lambda_2) =

- 6144 \lambda_2^{3}

- 80 \lambda_2 \lambda_3^{2}

- 16 \lambda_3^{3}

+ \frac{2816}{15} g_1^{2} \lambda_2^{2}

+ \frac{2816}{3} g_3^{2} \lambda_2^{2}

+ \frac{12}{5} g_1^{2} \lambda_3^{2}

+ 12 g_2^{2} \lambda_3^{2}

+ \frac{5732}{225} g_1^{4} \lambda_2

+ \frac{704}{45} g_1^{2} g_3^{2} \lambda_2

-  \frac{212}{3} g_3^{4} \lambda_2

+ \frac{4}{5} g_1^{4} \lambda_3

-  \frac{5416}{3375} g_1^{6}

-  \frac{2554}{675} g_1^{4} g_3^{2}

-  \frac{398}{135} g_1^{2} g_3^{4}

+ \frac{349}{27} g_3^{6}

- 448 \lambda_2^{2} \tr\left(L_t^{\dagger} L_t \right)

- 12 \lambda_3^{2} \tr\left(Y_u^{\dagger} Y_u \right)

- 12 \lambda_3^{2} \tr\left(Y_d^{\dagger} Y_d \right)

- 4 \lambda_3^{2} \tr\left(Y_e^{\dagger} Y_e \right)

+ \frac{16}{3} g_1^{2} \lambda_2 \tr\left(L_t^{\dagger} L_t \right)

+ \frac{80}{3} g_3^{2} \lambda_2 \tr\left(L_t^{\dagger} L_t \right)

-  \frac{32}{225} g_1^{4} \tr\left(L_t^{\dagger} L_t \right)

-  \frac{16}{45} g_1^{2} g_3^{2} \tr\left(L_t^{\dagger} L_t \right)

-  \frac{13}{18} g_3^{4} \tr\left(L_t^{\dagger} L_t \right)

- 12 \lambda_2 \tr\left(Y_u^{\dagger} Y_u L_t L_t^{\dagger} \right)

- 16 \lambda_2 \tr\left(L_t^{\dagger} L_t L_t^{\dagger} L_t \right)

+ 4 \tr\left(Y_u^{\dagger} Y_u L_t L_t^{\dagger} L_t L_t^{\dagger} \right)

+ 24 \tr\left(L_t^{\dagger} L_t L_t^{\dagger} L_t L_t^{\dagger} L_t \right),
\end{autobreak}
\end{align*}
\begin{align*}
\begin{autobreak}
\beta^{(1)}(\lambda_3) =

+ 12 \lambda_1 \lambda_3

+ 64 \lambda_2 \lambda_3

+ 8 \lambda_3^{2}

-  \frac{5}{2} g_1^{2} \lambda_3

-  \frac{9}{2} g_2^{2} \lambda_3

- 8 g_3^{2} \lambda_3

+ \frac{6}{25} g_1^{4}

+ 6 \lambda_3 \tr\left(Y_u^{\dagger} Y_u \right)

+ 6 \lambda_3 \tr\left(Y_d^{\dagger} Y_d \right)

+ 2 \lambda_3 \tr\left(Y_e^{\dagger} Y_e \right)

+ 4 \lambda_3 \tr\left(L_t^{\dagger} L_t \right)

- 4 \tr\left(Y_u^{\dagger} Y_u L_t L_t^{\dagger} \right),
\end{autobreak}
\end{align*}
\begin{align*}
\begin{autobreak}
\beta^{(2)}(\lambda_3) =

- 144 \lambda_1 \lambda_3^{2}

- 768 \lambda_2 \lambda_3^{2}

- 60 \lambda_1^{2} \lambda_3

- 1280 \lambda_2^{2} \lambda_3

- 52 \lambda_3^{3}

+ \frac{72}{5} g_1^{2} \lambda_1 \lambda_3

+ 72 g_2^{2} \lambda_1 \lambda_3

+ \frac{2048}{15} g_1^{2} \lambda_2 \lambda_3

+ \frac{2048}{3} g_3^{2} \lambda_2 \lambda_3

+ \frac{10}{3} g_1^{2} \lambda_3^{2}

+ 6 g_2^{2} \lambda_3^{2}

+ \frac{32}{3} g_3^{2} \lambda_3^{2}

+ \frac{12}{5} g_1^{4} \lambda

+ \frac{64}{5} g_1^{4} \lambda_2

+ \frac{57487}{3600} g_1^{4} \lambda_3

+ \frac{9}{8} g_1^{2} g_2^{2} \lambda_3

+ \frac{32}{9} g_1^{2} g_3^{2} \lambda_3

-  \frac{145}{16} g_2^{4} \lambda_3

- 44 g_3^{4} \lambda_3

-  \frac{2603}{750} g_1^{6}

-  \frac{9}{10} g_1^{4} g_2^{2}

-  \frac{8}{5} g_1^{4} g_3^{2}

- 72 \lambda_1 \lambda_3 \tr\left(Y_u^{\dagger} Y_u \right)

- 72 \lambda_1 \lambda_3 \tr\left(Y_d^{\dagger} Y_d \right)

- 24 \lambda_1 \lambda_3 \tr\left(Y_e^{\dagger} Y_e \right)

- 256 \lambda_2 \lambda_3 \tr\left(L_t^{\dagger} L_t \right)

- 24 \lambda_3^{2} \tr\left(Y_u^{\dagger} Y_u \right)

- 24 \lambda_3^{2} \tr\left(Y_d^{\dagger} Y_d \right)

- 8 \lambda_3^{2} \tr\left(Y_e^{\dagger} Y_e \right)

- 16 \lambda_3^{2} \tr\left(L_t^{\dagger} L_t \right)

+ \frac{17}{4} g_1^{2} \lambda_3 \tr\left(Y_u^{\dagger} Y_u \right)

+ \frac{5}{4} g_1^{2} \lambda_3 \tr\left(Y_d^{\dagger} Y_d \right)

+ \frac{15}{4} g_1^{2} \lambda_3 \tr\left(Y_e^{\dagger} Y_e \right)

+ \frac{8}{3} g_1^{2} \lambda_3 \tr\left(L_t^{\dagger} L_t \right)

+ \frac{45}{4} g_2^{2} \lambda_3 \tr\left(Y_u^{\dagger} Y_u \right)

+ \frac{45}{4} g_2^{2} \lambda_3 \tr\left(Y_d^{\dagger} Y_d \right)

+ \frac{15}{4} g_2^{2} \lambda_3 \tr\left(Y_e^{\dagger} Y_e \right)

+ 40 g_3^{2} \lambda_3 \tr\left(Y_u^{\dagger} Y_u \right)

+ 40 g_3^{2} \lambda_3 \tr\left(Y_d^{\dagger} Y_d \right)

+ \frac{40}{3} g_3^{2} \lambda_3 \tr\left(L_t^{\dagger} L_t \right)

-  \frac{38}{25} g_1^{4} \tr\left(Y_u^{\dagger} Y_u \right)

+ \frac{2}{5} g_1^{4} \tr\left(Y_d^{\dagger} Y_d \right)

- 2 g_1^{4} \tr\left(Y_e^{\dagger} Y_e \right)

-  \frac{4}{25} g_1^{4} \tr\left(L_t^{\dagger} L_t \right)

- 16 g_3^{4} \tr\left(Y_u^{\dagger} Y_u \right)

- 16 g_3^{4} \tr\left(Y_d^{\dagger} Y_d \right)

-  \frac{27}{2} \lambda_3 \tr\left(Y_u^{\dagger} Y_u Y_u^{\dagger} Y_u \right)

+ \lambda_3 \tr\left(Y_u^{\dagger} Y_u L_t L_t^{\dagger} \right)

- 21 \lambda_3 \tr\left(Y_u^{\dagger} Y_d Y_d^{\dagger} Y_u \right)

-  \frac{27}{2} \lambda_3 \tr\left(Y_d^{\dagger} Y_d Y_d^{\dagger} Y_d \right)

-  \frac{9}{2} \lambda_3 \tr\left(Y_e^{\dagger} Y_e Y_e^{\dagger} Y_e \right)

- 24 \lambda_3 \tr\left(L_t^{\dagger} L_t L_t^{\dagger} L_t \right)

-  \frac{8}{15} g_1^{2} \tr\left(Y_u^{\dagger} Y_u L_t L_t^{\dagger} \right)

-  \frac{32}{3} g_3^{2} \tr\left(Y_u^{\dagger} Y_u L_t L_t^{\dagger} \right)

+ 14 \tr\left(Y_u^{\dagger} Y_u Y_u^{\dagger} Y_u L_t L_t^{\dagger} \right)

+ 44 \tr\left(Y_u^{\dagger} Y_u L_t L_t^{\dagger} L_t L_t^{\dagger} \right)

- 2 \tr\left(Y_u^{\dagger} Y_d Y_d^{\dagger} Y_u L_t L_t^{\dagger} \right).
\end{autobreak}
\end{align*}

\bibliographystyle{JHEP}
\bibliography{main}

\end{document}